\documentclass[pra,aps,twocolumn,superscriptaddress,nofootinbib]{revtex4}
\usepackage{hyperref}
\usepackage{bm}
\usepackage{epsfig}
\usepackage{amssymb}
\usepackage{graphicx}
\usepackage{mathrsfs}
\usepackage{amsmath}

\begin{document}
\title{Gaussian intrinsic entanglement for states with partial minimum uncertainty}
\author{Ladislav Mi\v{s}ta, Jr. and Kl\'{a}ra Baksov\'{a}}
\affiliation{Department of Optics, Palack\' y University, 17.
listopadu 12,  771~46 Olomouc, Czech Republic}

\begin{abstract}
We develop a theory of a quantifier of bipartite Gaussian entanglement called Gaussian intrinsic entanglement (GIE)
which was proposed recently in [L. Mi\v{s}ta {\it et al.}, Phys. Rev. Lett. {\bf 117}, 240505 (2016)]. The GIE
provides a compromise between computable and physically meaningful entanglement quantifiers and so far
it was calculated for two-mode Gaussian states including all symmetric partial minimum-uncertainty states,
weakly mixed asymmetric squeezed thermal states with partial minimum uncertainty, and weakly mixed
symmetric squeezed thermal states. We improve the method of derivation of GIE and we show,
that all previously derived formulas for GIE of weakly mixed states in fact hold for states
with higher mixedness. In addition, we derive analytical formulas for GIE for several new
classes of two-mode Gaussian states with partial minimum uncertainty. Finally, it is shown,
that like for all previously known states, also for all new states the GIE is equal to
Gaussian R\'{e}nyi-2 entanglement of formation. This finding strengthens a
conjecture about equivalence of GIE and Gaussian R\'{e}nyi-2 entanglement
of formation for all bipartite Gaussian states.
\end{abstract}

\maketitle

\section{Introduction}\label{sec_Introduction}

Since quantum entanglement saw the light of day \cite{Einstein_35,Schrodinger_35}, it metamorphosed from a puzzling ingredient of quantum mechanics to a
unique concept opening new paradigms in communication and computing. The development of theory and experiment exploring entanglement unveiled,
that it is imperative not only to be able to certify its presence \cite{Peres_96,MHorodecki_96}, but also to quantify it properly.
For example, entanglement exhibits various monogamy properties \cite{Coffman_00,Koashi_04} which are quantitative, and therefore they
can be captured only with the help of entanglement measures. Likewise, entanglement measures are indispensable in proofs of impossibility \cite{Giedke_02,Eisert_02} or limitation \cite{Ohliger_10} of some quantum-information protocols, and they provide useful bounds on several important hardly computable quantities \cite{Rains_01,Christandl_07}. As far as the experiment is concerned, entanglement measures are needed to assess the quality of experimentally prepared entangled states \cite{Walborn_06} and entangling gates \cite{O'Brien_04}, and what is more, they are vital for verification of successful demonstration of some crucial concepts of quantum communication such as entanglement distillation \cite{Yamamoto_03} or the existence of a
gap between secret key content and distillable entanglement \cite{Dobek_11}.

Demand for entanglement quantifiers which would stay on solid grounds triggered the development of the axiomatic theory of entanglement measures \cite{Plenio_07,Vidal_00}. Primarily, any good entanglement measure should be a non-negative function, which vanishes on separable (disentangled)
states, and which does not increase under local operations and classical communication. Additionally, a good entanglement measure should be
also equal to marginal von Neumann entropy on pure states, it should be convex, additive on tensor product and asymptotically
continuous.

At present, there is a number of different entanglement measures which quantify
entanglement in different ways. The most widely used measure is undoubtedly a relatively
easily computable logarithmic negativity
\cite{Vidal_02,Eisert_PhD,Plenio_05}, which quantifies entanglement of a given quantum state
according to how much a partial transpose of the state deviates from a physical
state. Other means of entanglement quantification provide the so called operational measures known
as entanglement of formation and distillable entanglement \cite{Bennett_96}, which quantify how much
maximal pure-state entanglement one needs to create a shared quantum state and how much maximal pure-state
entanglement one can distill from a shared quantum state, respectively. A conceptually different way of entanglement
quantification provide geometric measures, which quantify the amount of
entanglement in a quantum state via a distance of the state from the
set of separable states \cite{Vedral_97}. Yet another way of entanglement
quantification exists, which utilizes information-theoretical measures of
correlations and the measure in question is the so called squashed entanglement
\cite{Christandl_04} defined as a quantum conditional
mutual information of an extension of the investigated
quantum state, which is minimized over all the extensions.

Each of the measures listed above has its advantages as well as weaknesses.
First, for most of the measures mentioned above, some of the axioms are relaxed.
Second, the measures either possess a good operational meaning or are computable but not both.
One exception to the first rule is the squashed entanglement, for which all axioms
imposed on a proper entanglement measure are satisfied \cite{Brandao_11}, but up to exceptions \cite{Christandl_05} it
is hard to compute. The other candidate for a good entanglement measure is the
entanglement of formation, which was so far computed for two qubits \cite{Wootters_98} and
symmetric Gaussian states \cite{Giedke_03a}, but which has operational meaning beyond
some practically utilizable task.

Recently, an attempt has been made \cite{Mista_15,Mista_16} to extend the family of candidates for a good
entanglement measure, aiming at the same time at probing the gap between computable and
physically meaningful entanglement measures. This resulted in the proposal of a new quantifier
of entanglement called intrinsic entanglement (IE) \cite{Mista_15,Mista_16}. The introduction of
the IE closely follows the idea of Gisin and Wolf \cite{Gisin_00} to quantify entanglement
by the amount of a resource for classical secret key agreement protocol \cite{Maurer_93},
the so called secret correlations, which one can extract by a measurement from the analyzed
state. Existing studies of IE \cite{Mista_15,Mista_16} focus on an important and experimentally
feasible Gaussian scenario, in which all states, channels and measurements are assumed to be Gaussian.
In this fully Gaussian framework it was shown, that the Gaussian IE (GIE) is faithful,
i.e., it vanishes if and only if the investigated state is
separable and it is monotonic under Gaussian local trace-preserving operations and classical communication.
Additionally, GIE was calculated for several classes of two-mode Gaussian states including all
pure states and symmetric states with a three-mode purification, as well as for ``weakly mixed''
asymmetric squeezed thermal states with a three-mode purification and symmetric squeezed thermal states.
Since in all the cases GIE is reached by feasible homodyne or heterodyne measurements, it represents an
experimentally accessible quantity. Besides, it was also shown that in all the cases GIE is equal
to another Gaussian entanglement measure know as Gaussian R\'{e}nyi-2 entanglement \cite{Adesso_12}
later dubbed more fittingly as Gaussian R\'{e}nyi-2 entanglement of formation (GR2EoF) \cite{Lami_17}.
This finding has led to the conjecture \cite{Mista_16} that the two quantities are equal on all bipartite
Gaussian states. Note, that the GR2EoF is a measure of Gaussian states, which is equipped with many
important properties. First, GR2EoF does not increase under all Gaussian local operations and classical communication,
and therefore it is a proper measure of Gaussian entanglement. Second, GR2EoF is additive on two-mode symmetric states
and can be interpreted in the context of the sampling entropy for Wigner quasiprobability distribution.
Third, as a cherry on the cake, GR2EoF satisfies \cite{Adesso_12} both monogamy inequality \cite{Coffman_00} as
well as Gaussian R\'{e}nyi-2 version of Koashi-Winter monogamy relation \cite{Koashi_04}.
Finally, GR2EoF is computable for all two-mode Gaussian states. To be more specific,
GR2EoF can be calculated analytically for all two-mode Gaussian states with a three-mode purification,
all symmetric states as well as two-mode squeezed thermal states, and numerically for all other two-mode
Gaussian states. If the conjecture proves to be true, all the properties of GR2EoF mentioned above
will be transferred to GIE and vice versa. As a consequence, we would have a unique experimentally meaningful
measure of Gaussian entanglement possessing all the properties listed above, which is computable on all two-mode
Gaussian states, and which is operationally associated to the secret key agreement protocol.

In this paper we further develop the theory of GIE and investigate its relation to GR2EoF. First, we show
that analytical formulas for GIE of symmetric two-mode squeezed thermal states and asymmetric squeezed thermal states with a
three-mode purification derived in Ref.~\cite{Mista_16} hold for a larger
set of the states. Next, we derive an analytical formula for GIE
for several new classes of two-mode Gaussian states with a three-mode purification.
Finally, we discuss the relation of GIE to other entanglement measures
encompassing logarithmic negativity and GR2EoF. It is shown, in particular, that the GIE for the new
classes of states is again equal to the GR2EoF, which further strengthens the conjectured equivalence of
the two quantities. As a byproduct of derivation of new formulas for GIE, we also obtain
an explicit form of a symplectic matrix which symplectically diagonalizes an arbitrary two-mode
covariance matrix in standard form the off-diagonal block of which has a negative determinant.


The paper is organized as follows. In Section~\ref{sec_IE} we
explain basics of a new quantifier of bipartite entanglement called intrinsic entanglement.
In Section~\ref{sec_Gaussian_states} we give a brief introduction into the formalism
of Gaussian states. Section~\ref{sec_GIE} is dedicated to the explanation of the
concept of Gaussian intrinsic entanglement. In Section~\ref{sec_U} and Section~\ref{sec_saturation}
we describe in detail a generic method of derivation of the Gaussian intrinsic entanglement. In Section~\ref{sec_GIE_for_GLEMS} we derive
Gaussian intrinsic entanglement for several new classes of two-mode Gaussian states with a three-mode purification. Section~\ref{sec_Relation_to_other_measures}
deals with relation of Gaussian intrinsic entanglement to logarithmic negativity and Gaussian R\'{e}nyi-2 entanglement of formation.
Finally, Section~\ref{sec_Conclusions} contains conclusions.

\section{Intrinsic entanglement}\label{sec_IE}

The definition of IE is based on the classical measure of entanglement \cite{Gisin_00} which
utilizes mapping of quantum states onto probability distributions. First, for the state of
interest $\rho_{AB}$ a purification $|\Psi\rangle_{ABE}$ is constructed, where
$\mbox{Tr}_{E}|\Psi\rangle_{ABE}\langle\Psi|=\rho_{AB}$. Next, the purification is mapped by
pure-state measurements $\{\Pi_{A}\}$ and $\{\Pi_{B}\}$, and a generic measurement $\{\Pi_{E}\}$, on
subsystems $A,B$ and $E$, onto a probability distribution
\begin{equation}\label{PABE}
P(A,B,E)=\mbox{Tr}(\Pi_{A}\otimes\Pi_{B}\otimes\Pi_{E}|\Psi\rangle_{ABE}\langle\Psi|).
\end{equation}
The key quantity in the definition of the classical measure of entanglement, which also stays behind
introduction of the squashed entanglement \cite{Christandl_04}, is the so called intrinsic
information \cite{Maurer_99} of the distribution,
\begin{eqnarray}\label{IntrinsicI}
I(A; B\downarrow E)=\mathop{\mbox{inf}}_{E\rightarrow
\tilde{E}}[I(A; B| \tilde{E})].
\end{eqnarray}
Here the infimum is taken over all conditional probability distributions
$P(\tilde{E}|E)$ defining a new random variable $\tilde{E}$, and
\begin{eqnarray}\label{conditionalI}
I(A; B|E)=H(A|E)-H(A|B,E),
\end{eqnarray}
is the mutual information between $A$ and $B$ conditioned on $E$.
Here, $H(X|Y)$ is the conditional Shannon entropy given by
$H(X|Y)=H(X,Y)-H(Y)$, where $H(X,Y)$ and $H(Y)$ are joint
and marginal Shannon entropies \cite{Shannon_48}, respectively.

From Eqs.~(\ref{IntrinsicI}) and (\ref{conditionalI}) it follows
that the intrinsic information quantifies how much reduces Bob's
uncertainty about Alice's variable $A$ if he looks at his variable
$B$ after Eve announces her variable $E$ (or a function of her
variable) \cite{Dusek_06}. The intrinsic information also provides
an upper bound \cite{Maurer_99} (not always tight
\cite{Renner_03}) on the rate at which a secret key can be
generated from the probability distribution $P$ in the secret key
agreement protocol \cite{Maurer_93}, and more importantly, it is
conjectured to be equal to a secret key rate in the modification
of the protocol called public Eve scenario
\cite{Christandl_04b,PES}.

Because Alice and Bob may perform unsuitable measurements
on an entangled state such, that intrinsic information vanishes, and on the other hand,
a bad measurement on Eve's side may allow Alice and Bob to get
a strictly positive intrinsic information even for a separable state \cite{Gisin_00},
some optimization is needed to get a quantity which faithfully maps entanglement onto
secret correlations. For this reason, Gisin and Wolf defined the classical measure of
entanglement as the following optimized intrinsic information \cite{Gisin_00}:
\begin{eqnarray}\label{mu}
\mu(\rho_{AB})=\mathop{\mbox{inf}}_{\left\{\Pi_{E},|\Psi\rangle\right\}}
\left\{\mathop{\mbox{sup}}_{\left\{\Pi_{A},\Pi_{B}\right\}}
\left[I\left(A;B\downarrow E\right)\right]\right\},
\end{eqnarray}
where the minimization is carried out also over all purifications of the studied
state $\rho_{AB}$. The IE is then obtained \cite{Mista_15,Mista_16} by reversing the order of
optimization in the previous formula, i.e.,
\begin{equation}\label{Edownarrow}
E_{\downarrow}(\rho_{AB})=\mathop{\mbox{sup}}_{\left\{\Pi_{A},\Pi_{B}\right\}}
\left\{\mathop{\mbox{inf}}_{\left\{\Pi_{E},|\Psi\rangle\right\}}\left[I\left(A;
B\downarrow E\right)\right]\right\}.
\end{equation}

In the rest of the present paper we investigate the IE for the case, when all states $\rho_{AB}$,
purifications $|\Psi\rangle_{ABE}$, measurements $\Pi_{A}, \Pi_{B}$ and $\Pi_{E}$, as well as the
conditional probability distributions $P(\tilde{E}|E)$, are Gaussian. Therefore, in the following
section we give a brief introduction into the theory of Gaussian quantum states.
\section{GAUSSIAN STATES}\label{sec_Gaussian_states}

In this paper we work with quantum states of systems with
infinite-dimensional Hilbert state space, which we shall call
modes in what follows. A system of $N$ modes is described by a
vector of quadratures $\xi=(x_1,p_1,\ldots,x_N,p_N)^{T}$ with
components satisfying the canonical commutation rules
$[\xi_j,\xi_k]=i(\Omega_N)_{jk}$, where
\begin{equation}\label{OmegaN}
\Omega_{N}=\bigoplus_{i=1}^{N}J,\quad J=\left(\begin{array}{cc}
0 & 1 \\
-1 & 0\\
\end{array}\right),
\end{equation}
is the so called symplectic matrix. We restrict our attention to Gaussian states of modes, which
are defined as states with a Gaussian Wigner function. Any $N$-mode Gaussian state is thus fully characterized
by a $2N\times 2N$ covariance matrix (CM) $\gamma$ with entries $\gamma_{jk}=\langle\xi_{j}\xi_k+\xi_k\xi_j\rangle-2\langle\xi_j\rangle\langle\xi_k\rangle$
and by a $2N\times 1$ vector of first moments $\langle\xi\rangle$. Since the first moments
can be set to zero by local displacements which do not change entanglement of the state, they are irrelevant
as far as the entanglement is concerned and therefore they are assumed to be zero in the rest of the paper.

Apart from Gaussian states we will also utilize Gaussian unitary
operations defined as unitary operations which preserve Gaussian
character of states. On the CM level an $N$-mode Gaussian unitary
is represented by a real $2N\times2N$ symplectic matrix $S$
satisfying condition
\begin{equation}\label{Scondition}
S\Omega_{N} S^{T}=\Omega_{N}
\end{equation}
and it transforms a given CM $\gamma$ to $\gamma'=S\gamma S^{T}$.

In this paper we focus on Gaussian states of two-modes $A$ and
$B$, and therefore we denote CMs of the states as $\gamma_{AB}$.
The quantifiers of Gaussian entanglement including GIE \cite{Mista_15}
are always invariant with respect to local Gaussian unitary operations on modes $A$ and $B$,
and thus we can work without loosing any generality only with a
canonical form of the CM with respect to the operations, the so
called standard form \cite{Simon_00},
\begin{eqnarray}\label{gammast1}
\gamma_{AB}=\left(\begin{array}{cccc}
a & 0 & c_x & 0\\
0 & a & 0 & c_p \\
c_x & 0 & b & 0 \\
0 & c_p & 0 & b \\
\end{array}\right),
\end{eqnarray}
where we can assume $c_{x}\geq|c_{p}|\geq0$.
Since states with $c_{x}c_{p}\geq0$ are separable \cite{Simon_00}
and thus possess zero GIE \cite{Mista_15}, in calculations we can
restrict ourself only to CMs satisfying $c_{x}c_{p}<0$.
Introducing new more convenient parameters $k_{x}\equiv c_{x}$ and
$k_{p}\equiv|c_{p}|=-c_{p}$, we arrive at the following standard-form CM
which we shall consider in what follows \cite{Giedke_03a}:
\begin{eqnarray}\label{gammast2}
\gamma_{AB}=\left(\begin{array}{cccc}
a & 0 & k_x & 0\\
0 & a & 0 & -k_p \\
k_x & 0 & b & 0 \\
0 & -k_p & 0 & b \\
\end{array}\right),
\end{eqnarray}
where $k_x\geq k_p>0$. Construction of explicit examples of CMs of entangled Gaussian states
requires to know when the matrix (\ref{gammast2}) corresponds to a physical quantum state and when
the state is entangled. For this purpose, we can use a necessary and sufficient condition for a strictly
positive matrix (\ref{gammast2}) to be a CM of a physical quantum state, which is given by the following
inequalities \cite{Giedke_01}:
\begin{eqnarray}\label{CMcondition}
(ab-k_x^2)(ab-k_p^2)+1&\geq& a^2+b^2-2k_xk_p,\nonumber\\
ab-k_x^2&\geq& 1.
\end{eqnarray}
Additionally, CM (\ref{gammast2}) describes an entangled state if and only if \cite{Giedke_01}
\begin{eqnarray}\label{entcondition}
(ab-k_x^2)(ab-k_p^2)+1<a^2+b^2+2k_xk_p.
\end{eqnarray}

Further, we also need to find a Gaussian purification of the state
with CM (\ref{gammast2}), i.e., a pure Gaussian state
$|\Psi\rangle_{ABE}$ satisfying condition
$\mbox{Tr}_{E}|\Psi\rangle_{ABE}\langle\Psi|=\rho_{AB}$. This can
be done easily with the help of Williamson theorem
\cite{Williamson_36} which says, that for any two-mode CM
$\gamma_{AB}$ there is always a symplectic transformation $S$
which brings the CM to the normal form,
\begin{equation}\label{Williamson}
S\gamma_{AB}S^{T}=\mbox{diag}(\nu_{1},\nu_{1},\nu_{2},\nu_{2})\equiv\gamma_{AB}^{(0)},
\end{equation}
where $\nu_{1}\geq\nu_{2}\geq1$ are symplectic eigenvalues of
$\gamma_{AB}$. In physical terms Williamson theorem tells us that
for any two-mode Gaussian state there always exists a global
Gaussian unitary which brings the state into a tensor product of
two thermal states with CMs $\nu_{j}\openone$, $j=1,2$, where $\openone$ is
the $2\times2$ identity matrix.

The symplectic eigenvalues of CM (\ref{gammast2}) can be calculated
conveniently from the eigenvalues of the matrix $i\Omega\gamma_{AB}$
which are of the form $\{\pm \nu_{1},\pm \nu_2\}$ \cite{Vidal_02}.
In terms of parameters $a,b,k_{x}$ and $k_{p}$ they read explicitly as
\begin{equation}\label{nu12}
\nu_{1,2}=\sqrt{\frac{\Delta\pm\sqrt{D}}{2}},
\end{equation}
where
\begin{eqnarray}\label{DeltaD}
\Delta&=&a^2+b^2-2k_xk_p,\nonumber\\
D&=&\Delta^2-4\mbox{det}\gamma_{AB}=\left(a^2-b^2\right)^2+4M\tilde{M}
\end{eqnarray}
with
\begin{equation}\label{MtM}
M\equiv ak_x-bk_p,\quad \tilde{M}\equiv bk_x-ak_p.
\end{equation}

Similarly, we can express the symplectic matrix $S$ which brings
the CM to Williamson normal form (\ref{Williamson}) in terms of
parameters $a,b,k_{x}$ and $k_{p}$. This can be done using either
a method of Ref.~\cite{Serafini_05} or a method of
Ref.~\cite{Pirandola_09}. The derivation of the matrix for an
arbitrary CM (\ref{gammast2}) by means of the method of
Ref.~\cite{Serafini_05} is rather technical and it is placed into
the Appendix \ref{Sdiagonalization}.

Having now both symplectic eigenvalues and symplectic matrix from
Eq.~(\ref{Williamson}) in hands, we can proceed to the
construction of a CM $(\equiv \gamma_{\pi})$ of a purification of
the state with CM (\ref{gammast2}). Obviously, the structure of
the purification will depend on the so called symplectic rank $R$ of the CM,
which is defined as the number of its symplectic eigenvalues different from 1 \cite{Adesso_06b}.

In the most simple case of $R=0$ CM (\ref{gammast2}) describes a
pure state $|\psi\rangle_{AB}$ and the purifying subsystem $E$ is
completely independent of modes $A$ and $B$. Consequently,
$|\Psi\rangle_{ABE}=|\psi\rangle_{AB}|\varphi\rangle_{E}$, where
$|\varphi\rangle_{E}$ is the state of a purifying system, and thus
$\gamma_{\pi}=\gamma_{AB}\oplus\gamma_{E}$, where $\gamma_{E}$ is
a CM of the state $|\varphi\rangle_{E}$.

For $R>0$ the construction of the purification relies on
replacement of each of the $R$ modes with symplectic eigenvalue
$\nu_{i}>1$, $i=1,\ldots,R$, in the Williamson normal form
(\ref{Williamson}),
\begin{eqnarray}\label{gammaAB0}
\gamma_{AB}^{(0)}\equiv\left(\bigoplus_{i=1}^{R}\nu_{i}\openone\right)\bigoplus\openone_{2(2-R)},
\end{eqnarray}
where $\openone_{K}$ is the $K\times K$ identity matrix, with one
mode of the two-mode squeezed vacuum state with CM
\begin{equation}\label{TMSV}
\gamma_{\rm TMSV}(\nu_{i})=\left(\begin{array}{cc}
\nu_{i}\openone & \sqrt{\nu_{i}^{2}-1}\sigma_{z}\\
\sqrt{\nu_{i}^{2}-1}\sigma_{z} & \nu_{i}\openone\\
\end{array}\right).
\end{equation}
Hence, we get the following $(2+R)$-mode CM
\begin{equation}\label{gammapi0}
\gamma_{\pi}^{(0)}=\left(\begin{array}{cc}
\gamma_{AB}^{(0)} & \gamma_{ABE}^{(0)}\\
(\gamma_{ABE}^{(0)})^{T} & \gamma_{E}^{(0)}\\
\end{array}\right),\quad
\end{equation}
with
\begin{eqnarray}\label{gammaABEgammaE0}
\gamma_{ABE}^{(0)}=\left(\begin{array}{c}
\bigoplus_{i=1}^{R}\sqrt{\nu_{i}^{2}-1}\sigma_{z} \\
\mathbb{O}_{2(2-R)\times2R} \\
\end{array}\right),\quad\gamma_{E}^{(0)}=\bigoplus_{i=1}^{R}\nu_{i}\openone,
\end{eqnarray}
where $\sigma_{z}=\mbox{diag}(1,-1)$ is the diagonal Pauli-$z$
matrix and $\mathbb{O}_{I\times J}$ is the $I\times J$ zero matrix.

Next, we apply a symplectic matrix $S^{-1}\oplus\openone_{E}$,
where matrix $S$ brings CM (\ref{gammast2}) to the Williamson
normal form and $\openone_{E}$ is the $2R\times 2R$ identity
matrix, to CM (\ref{gammapi0}) which gives the sought CM of the
purification
\begin{eqnarray}\label{gammapi}
\gamma_{\pi}=\left(\begin{array}{cc}
\gamma_{AB} & \gamma_{ABE} \\
\gamma_{ABE}^{T} & \gamma_{E} \\
\end{array}\right),
\end{eqnarray}
where
\begin{eqnarray}\label{gammaABEgammaE}
\gamma_{ABE}=S^{-1}\gamma_{ABE}^{(0)},\quad\gamma_{E}=\gamma_{E}^{(0)}
\end{eqnarray}
and where we have used equality
\begin{eqnarray}\label{gammaAB}
\gamma_{AB}=S^{-1}\gamma_{AB}^{(0)}(S^{T})^{-1}.
\end{eqnarray}

\section{GAUSSIAN INTRINSIC ENTANGLEMENT}\label{sec_GIE}

Having established all needed ingredients we are now in a position
to provide a definition of GIE and give analytical formulas for
it, which have been obtained so far. In the case of a two-mode
Gaussian state $\rho_{AB}$ the GIE is defined as \cite{Mista_15}
\begin{equation}\label{GIE}
E_{\downarrow}^{G}\left(\rho_{AB}\right)=\mathop{\mbox{sup}}_{\Gamma_{A},\Gamma_{B}}
\mathop{\mbox{inf}}_{\Gamma_{E}}[I(A;B|E)],
\end{equation}
where
\begin{equation}\label{IABE}
I(A;B|E)=\frac{1}{2}\ln\left(\frac{\mbox{det}\sigma_{A}\mbox{det}\sigma_{B}}{\mbox{det}\sigma_{AB}}\right).
\end{equation}
Here
\begin{eqnarray}\label{sigma}
\sigma_{AB}&=&\gamma_{AB|E}+\Gamma_{A}\oplus\Gamma_{B},
\end{eqnarray}
where $\sigma_{A,B}$ are local submatrices of $\sigma_{AB}$ and
$\Gamma_{A}$ and $\Gamma_{B}$ are single-mode CMs of pure-state Gaussian
measurements on modes $A$ and $B$, respectively. Further,
\begin{eqnarray}\label{gammacond}
\gamma_{AB|E}&=&\gamma_{AB}-\gamma_{ABE}(\gamma_{E}+\Gamma_{E})^{-1}\gamma_{ABE}^{T}
\end{eqnarray}
is a CM of a conditional state $\rho_{AB|E}$ \cite{Giedke_02} of modes
$A$ and $B$ obtained by a Gaussian measurement with CM $\Gamma_E$ on
purifying subsystem $E$ of the purification with CM (\ref{gammapi}).
The use of Eq.~(\ref{gammaABEgammaE}) on the right-hand side (RHS) of
Eq.~(\ref{gammacond}) further yields for the CM $\gamma_{AB|E}$
the expression
\begin{equation}\label{gammacondS}
\gamma_{AB|E}=S^{-1}\gamma_{AB|E}^{(0)}(S^{-1})^{T}
\end{equation}
with
\begin{eqnarray}\label{gammacond0}
\gamma_{AB|E}^{(0)}&=&\gamma_{AB}^{(0)}-\gamma_{ABE}^{(0)}(\gamma_{E}^{(0)}+\Gamma_{E})^{-1}(\gamma_{ABE}^{(0)})^{T},
\end{eqnarray}
where matrices $\gamma_{AB}^{(0)}, \gamma_{ABE}^{(0)}$  and $\gamma_{E}^{(0)}$ are given
in Eqs.~(\ref{gammaAB0}) and (\ref{gammaABEgammaE0}).

Before summarizing currently known formulas for GIE, let us make a brief remark on uniqueness of the definition of GIE.
Namely, it is obvious that the symplectic transformation $S$ that brings CM $\gamma_{AB}$ to Williamson normal
form, Eq.~(\ref{Williamson}), is not determined uniquely. More precisely, if CM $\gamma_{AB}$ has non-degenerate
(degenerate) symplectic eigenvalues, $\nu_{1}\ne\nu_{2}$ ($\nu_{1}=\nu_{2}$), $S$ is determined uniquely up to
local orthogonal symplectic transformations $O_{A}$ and $O_{B}$ (global orthogonal symplectic transformation
$O_{AB}$) on modes $A$ and $B$ \cite{Serafini_05}. Nevertheless, despite the ambiguity in determination
of matrix $S$, the GIE is determined uniquely. To show this, imagine that instead of using symplectic matrix $S$,
we would use in CM of purification (\ref{gammapi}) a symplectic matrix $\bar{S}=OS$ ,
where $O=O_{A}\oplus O_{B}$ ($O=O_{AB}$) in the non-degenerate (degenerate) case. The change $S\rightarrow \bar{S}$ entails,
that the correlation matrix (\ref{sigma}) changes to
\begin{eqnarray}\label{barsigma}
\bar{\sigma}_{AB}&=&\bar{\gamma}_{AB|E}+\Gamma_{A}\oplus\Gamma_{B},
\end{eqnarray}
where
\begin{equation}\label{gammacondtildeS}
\bar{\gamma}_{AB|E}=S^{-1}O^{T}\gamma_{AB|E}^{(0)}O(S^{-1})^{T}
\end{equation}
and $\gamma_{AB|E}^{(0)}$ is given in Eq.~(\ref{gammacond0}). Further, making use in the latter formula
Eqs.~(\ref{Williamson}) and (\ref{gammaABEgammaE0}), and utilizing orthogonality of matrix $O$, one finds that
CM (\ref{gammacondtildeS}) boils down to
\begin{equation}\label{gammacondtildeS2}
\bar{\gamma}_{AB|E}=S^{-1}\bar{\gamma}_{AB|E}^{(0)}(S^{-1})^{T},
\end{equation}
where
\begin{equation}\label{bargammaABEcond0}
\bar{\gamma}_{AB|E}^{(0)}\equiv\gamma_{AB}^{(0)}-\gamma_{ABE}^{(0)}(\gamma_{E}^{(0)}+\bar{\Gamma}_{E})^{-1}(\gamma_{ABE}^{(0)})^{T}.
\end{equation}
Here $\bar{\Gamma}_{E}=\bar{O}^{T}\Gamma_{E}\bar{O}$ is a CM of a new Gaussian measurement with
\begin{equation}\label{tildeO}
\bar{O}=\left\{\begin{array}{lll} \sigma_{z}O_{A}\sigma_{z}, & \textrm{if} & R=1;\\
(\sigma_{z}\oplus\sigma_{z})O(\sigma_{z}\oplus\sigma_{z}), & \textrm{if} & R=2 ,\\
\end{array}\right.
\end{equation}
being an orthogonal symplectic matrix with $O=O_{A}\oplus O_{B}$ if $\nu_{1}\ne\nu_{2}$ and $O=O_{AB}$ if $\nu_{1}=\nu_{2}$.
Hence we see, that if we use for calculation of GIE symplectic matrix $\bar{S}$ instead of symplectic matrix $S$, the conditional mutual information (\ref{IABE}) that is to be optimized is obtained by replacing correlation matrix $\sigma_{AB}$ on the RHS of Eq.~(\ref{IABE}) with correlation matrix (\ref{barsigma}), which is further equivalent to calculation with the original correlation matrix $\sigma_{AB}$, Eq.~(\ref{sigma}), in which CM $\Gamma_{E}$
is replaced with CM $\bar{\Gamma}_{E}$. Since in the definition of GIE (\ref{GIE})
we carry out minimization over all CMs $\Gamma_{E}$, also the new CM $\bar{\Gamma}_{E}$ runs over all CMs in the course of the
minimization. Consequently, minimization with respect to all CMs $\Gamma_{E}$ of the conditional mutual information calculated from
correlation matrix $\bar{\sigma}_{AB}$ can be replaced with minimization over all CMs $\bar{\Gamma}_{E}$ and thus
we get the same value of GIE irrespective of whether we work with symplectic matix $S$ or $\bar{S}$, as we set out to prove.

Up to now, GIE was calculated for the following three classes of
states with CM (\ref{gammast2}) \cite{Mista_15,Mista_16}:

(1) {\it Symmetric GLEMS}. The GLEMS are Gaussian states with least negativity for given global and local purities
\cite{Adesso_04,Adesso_04b}. Entangled GLEMS satisfy equality $\nu_{2}=1$ and if they are symmetric they also fulfill condition $a=b$.
For all symmetric entangled GLEMS ($\equiv\rho_{AB}^{(1)}$) GIE reads as
\begin{equation}\label{GIEsymGLEMS}
E_{\downarrow}^{G}\left(\rho_{AB}^{(1)}\right)=\ln\left(\frac{a}{\sqrt{a^2-k_p^2}}\right).
\end{equation}

Further, if $k_{x}=k_{p}\equiv k$, symmetric GLEMS satisfy
$a^2-k^2=1$, they reduce to pure states ($\equiv\rho^{(p)}_{AB}$), and
the GIE is given by \cite{Mista_15}
\begin{equation}\label{GIEpure}
E_{\downarrow}^{G}\left(\rho_{AB}^{(p)}\right)=\ln(a).
\end{equation}

(2) {\it Symmetric squeezed thermal states} \cite{Botero_03}. The
squeezed thermal states are characterized by the condition
$k_{x}=k_{p}\equiv k$. The symmetric squeezed thermal states
($\equiv\rho_{AB}^{(2)}$) further fulfill condition $a=b$, and
they are entangled iff $a-k<1$ \cite{Simon_00,Giedke_03a}. For all
entangled symmetric squeezed thermal states satisfying inequality
$a\leq2.41$ GIE is equal to
\begin{equation}\label{GIEsymTMST}
E_{\downarrow}^{G}\left(\rho_{AB}^{(2)}\right)=\ln\left[\frac{(a-k)^2+1}{2(a-k)}\right].
\end{equation}

(3) {\it Asymmetric squeezed thermal GLEMS}. The states ($\equiv\rho_{AB}^{(3)}$) satisfy
conditions $k_{x}=k_{p}\equiv k$ and $\nu_{2}=1$, whereas $a$ and
$b$ generally differ. For all the states for which
$\sqrt{ab}\leq2.41$ GIE is given by
\begin{equation}\label{GIEasymGLEMS}
E_{\downarrow}^{G}\left(\rho_{AB}^{(3)}\right)=\ln\left(\frac{a+b}{|a-b|+2}\right).
\end{equation}

Previous analytical expressions for GIE can be derived in two steps.
The first step consists of calculation of an easier computable
upper bound on GIE. In the second step it is shown, that for some fixed
measurements on modes $A$ and $B$, the minimum of the conditional mutual
information (\ref{IABE}) over all measurements on subsystem $E$ saturates
the bound. As a consequence, the upper bound is tight and it gives the
sought value of the GIE. The derivation of the upper
bound for symmetric squeezed thermal states and asymmetric squeezed thermal
GLEMS proved to be tractable only for ``weakly mixed'' states satisfying
inequalities $a\leq2.41$ and $\sqrt{ab}\leq2.41$, respectively, which is
the cause why formulas (\ref{GIEsymTMST}) and (\ref{GIEasymGLEMS})
are currently known to hold only for states fulfilling the latter
inequalities.

In the next section we improve the method of derivation of
GIE, which is later used for calculation of GIE for new classes of
two-mode Gaussian states. As a byproduct, we get a stronger condition
under which formulas (\ref{GIEsymTMST}) and (\ref{GIEasymGLEMS}) are valid
thus extending the set of states for which GIE is known.

\section{UPPER BOUND ON GIE}\label{sec_U}

A key role in derivation of analytical formulas for GIE given in
Eqs.~(\ref{GIEsymGLEMS}), (\ref{GIEpure}), (\ref{GIEsymTMST}) and
(\ref{GIEasymGLEMS}) plays the so called Gaussian classical mutual
information (GCMI) of a bipartite Gaussian quantum state. This
quantity has been introduced in Ref.~\cite{Mista_11} by
restricting the classical mutual information of a quantum state
$\rho_{AB}$ \cite{DiVincenzo_04},
\begin{equation}\label{I}
\mathcal{I}_{c}\left(\rho_{AB}\right)\equiv\mathop{\mbox{sup}}_{\Pi_{A},\Pi_{B}}[I(A;B)],
\end{equation}
to Gaussian states and measurements. Here $I(A;B)=H(A)+H(B)-H(A,B)$ is the classical mutual information of
the probability distribution $P(A,B)=\mbox{Tr}[(\Pi_{A}\otimes\Pi_B)\rho_{AB}]$ of
outcomes of local measurements $\Pi_{A}$ and $\Pi_B$ on state $\rho_{AB}$. In this way, one gets for a
Gaussian state $\rho_{AB}$ with CM $\gamma_{AB}$ the GCMI in the
form \cite{Mista_11}:
\begin{equation}\label{IcGrhoAB}
\mathcal{I}_{c}^{G}\left(\rho_{AB}\right)=\mathop{\mbox{sup}}_{\Gamma_{A},\Gamma_{B}}[I(A;B)],
\end{equation}
with
\begin{equation}\label{IAB}
I(A;B)=\frac{1}{2}\ln\left[\frac{\mbox{det}(\gamma_{A}+\Gamma_{A})\mbox{det}(\gamma_{B}+\Gamma_{B})}
{\mbox{det}(\gamma_{AB}+\Gamma_{A}\oplus\Gamma_{B})}\right],
\end{equation}
where $\gamma_{A,B}$ are local CMs of $\gamma_{AB}$.

For a generic two-mode Gaussian state with the standard form CM
(\ref{gammast1}) optimization in Eq.~(\ref{IcGrhoAB}) requires
finding of roots of a 12th-order polynomial \cite{Mista_11}, which
can be done generally only numerically. Nevertheless, for a
certain region of parameters $a, b, c_{x}$ and $c_{p}$ of the CM,
the optimization can be performed analytically. Specifically, if
the parameters satisfy inequality \cite{Mista_16}
\begin{equation}\label{G}
G\equiv\sqrt{\frac{a}{b}}+\sqrt{\frac{b}{a}}+\frac{1}{\sqrt{ab}}-\sqrt{ab-c_{x}^2}\geq0,
\end{equation}
the GCMI reads as
\begin{equation}\label{IcGhom}
\mathcal{I}_{c,h}^{G}\left(\rho_{AB}\right)=\frac{1}{2}\ln\left(\frac{ab}{ab-c_{x}^{2}}\right),
\end{equation}
and it is reached by double homodyne detection of quadratures
$x_{A}$ and $x_{B}$ on modes $A$ and $B$. Needles to say for
completeness, that if an opposite inequality $G<0$ holds, then
homodyning is not optimal anymore, and a larger value of GCMI can
be obtained, e.g., by double heterodyne detection on modes $A$ and
$B$, i.e., projection of the modes onto coherent states.

Previous findings about GCMI represent a backbone of the method
used in Refs.~\cite{Mista_15,Mista_16} to evaluate formulas
(\ref{GIEsymGLEMS})--(\ref{GIEasymGLEMS}) for GIE for various
classes of two-mode Gaussian states. The method consists of
calculation of an easier computable quantity,
\begin{equation}\label{Ubound}
U\left(\rho_{AB}\right)\equiv\mathop{\mbox{inf}}_{\Gamma_{E}}\left[\mathcal{I}_{c}^{G}\left(\rho_{AB|E}\right)\right],
\end{equation}
where
\begin{equation}\label{IcG}
\mathcal{I}_{c}^{G}\left(\rho_{AB|E}\right)=\mathop{\mbox{sup}}_{\Gamma_{A},\Gamma_{B}}[I(A;B|E)]
\end{equation}
is the GCMI of the conditional state $\rho_{AB|E}$ with CM
(\ref{gammacondS}), which is an upper bound on GIE as follows from
the max-min inequality \cite{Boyd_04},
$E_{\downarrow}^{G}(\rho_{AB})\leq U(\rho_{AB})$. In the next
step, for some fixed CMs $\Gamma_{A}'$ and $\Gamma_{B}'$ we find
$\mathop{\mbox{inf}}_{\Gamma_{E}}[I(A;B|E)]$,
which saturates the bound, and thus the bound
gives us the value of GIE we are looking for,
\begin{equation}\label{GIEeqU}
E_{\downarrow}^{G}(\rho_{AB})=U(\rho_{AB}).
\end{equation}

A specific feature of all states for which GIE was calculated so
far, i.e., states for which Eqs.~(\ref{GIEsymGLEMS})--(\ref{GIEasymGLEMS})
hold, is that for any CM $\Gamma_{E}$ the optimal measurement on modes $A$ and $B$
of the conditional state $\rho_{AB|E}$ in the standard form,
which reaches GCMI (\ref{IcG}), is always homodyne detection of quadratures
$x_{A}$ and $x_{B}$. This property in fact makes evaluation of GIE possible and
for this reason, in the present paper we will also restrict ourself to states
equipped with this property.

To find the condition under which for a given Gaussian state the GCMI
(\ref{IcG}) is attained by double homodyning for any choice of CM
$\Gamma_{E}$, one can use inequality (\ref{G}). Since the GCMI
is invariant with respect to local symplectic transformations, the CM
of the conditional state $\rho_{AB|E}$ can be taken in the standard form
\begin{eqnarray}\label{gammacondst}
\gamma_{AB|E}=\left(\begin{array}{cccc}
\tilde{a} & 0 & \tilde{c}_x & 0\\
0 & \tilde{a} & 0 & \tilde{c}_p \\
\tilde{c}_x & 0 & \tilde{b} & 0 \\
0 & \tilde{c}_p & 0 & \tilde{b} \\
\end{array}\right),
\end{eqnarray}
where $\tilde{c}_x\geq|\tilde{c}_p|\geq0$. In terms of the
parameters of CM (\ref{gammacondst}) condition (\ref{G}) reads as
\begin{equation}\label{tildeG}
\tilde{G}=\sqrt{\frac{\tilde{a}}{\tilde{b}}}+\sqrt{\frac{\tilde{b}}{\tilde{a}}}+\frac{1}{\sqrt{\tilde{a}\tilde{b}}}-\sqrt{\tilde{a}\tilde{b}-\tilde{c}_{x}^2}\geq0.
\end{equation}
In order for a given Gaussian state $\rho_{AB}$ the GCMI
(\ref{IcGrhoAB}) to be reached by double homodyning, i.e., to be
of the form
\begin{equation}\label{IcGcondhom}
\mathcal{I}_{c,h}^{G}\left(\rho_{AB|E}\right)=\frac{1}{2}\ln\left(\frac{\tilde{a}\tilde{b}}{\tilde{a}\tilde{b}-\tilde{c}_{x}^{2}}\right),
\end{equation}
for any CM $\Gamma_{E}$, inequality (\ref{tildeG}) has to be
fulfilled for any CM $\Gamma_{E}$. In Ref.~\cite{Mista_16} it was
shown, that if a symmetric squeezed thermal state satisfies
inequality $a\leq 2.41$, then inequality (\ref{tildeG}) holds for
any CM $\Gamma_{E}$. Similarly, in Ref.~\cite{Mista_16} the
validity of inequality (\ref{tildeG}) for any CM $\Gamma_{E}$ was
also shown for asymmetric squeezed thermal GLEMS fulfilling
condition $\sqrt{ab}\leq2.41$. In what follows, we derive a
stronger condition under which inequality (\ref{tildeG}) is
satisfied for any CM $\Gamma_{E}$ thereby extending the set of
states for which GIE is known.

We start with an observation \cite{Lami_17,Horn_85}, that the CM
$\gamma_{AB}$ of the investigated state $\rho_{AB}$, and
the CM $\gamma_{AB|E}$ of the conditional state $\rho_{AB|E}$,
satisfy inequality $\gamma_{AB}\geq\gamma_{AB|E}$.
Further, both matrices appearing in the
latter inequality are physical CMs which are positive definite
\cite{Simon_00,Serafini_07}, which together with the latter inequality
implies that $\mbox{det}\gamma_{AB}\geq\mbox{det}\gamma_{AB|E}$
\cite{Horn_85}. As a consequence, the following inequality holds:
\begin{equation}\label{nu12ineq1}
\nu_{1}^{2}\nu_{2}^{2}\geq
(\tilde{a}\tilde{b}-\tilde{c}_{x}^{2})(\tilde{a}\tilde{b}-\tilde{c}_{p}^{2})\geq(\tilde{a}\tilde{b}-\tilde{c}_{x}^{2})^{2}.
\end{equation}
Here, to get the first inequality we used Eqs.~(\ref{Williamson})
and (\ref{gammacondst}), whereas the second inequality follows
from inequality $\tilde{c}_{x}\geq|\tilde{c}_{p}|$. By taking
finally the fourth root of inequality $\nu_{1}^{2}\nu_{2}^{2}\geq
(\tilde{a}\tilde{b}-\tilde{c}_{x}^{2})^{2}$, we obtain
\begin{equation}\label{nu12ineq}
\sqrt{\nu_{1}\nu_{2}}\geq
\sqrt{\tilde{a}\tilde{b}-\tilde{c}_{x}^{2}}.
\end{equation}

Matrix inequality $\gamma_{AB}\geq\gamma_{AB|E}$ also imposes a
restriction on local symplectic eigenvalues $\tilde{a}$ and
$\tilde{b}$ appearing in the standard form of CM $\gamma_{AB|E}$,
Eq.~(\ref{gammacondst}). Consider now the CM $\gamma_{AB|E}$ of
the conditional state $\rho_{AB|E}$ after a Gaussian measurement
with a generic CM $\Gamma_{E}$ on a purifying subsystem $E$ of the
state $\rho_{AB}$, expressed with respect to $A|B$ splitting,
\begin{eqnarray}\label{gammacondblock}
\gamma_{AB|E}=\left(\begin{array}{cc}
\tilde{A} & \tilde{C} \\
\tilde{C}^{T} & \tilde{B} \\
\end{array}\right),
\end{eqnarray}
which will not be generally in the standard form
(\ref{gammacondst}). Inequality $\gamma_{AB}\geq\gamma_{AB|E}$
then implies the following inequalities for the local CMs of modes
$A$ and $B$, $a\openone\geq\tilde{A}$, and
$b\openone\geq\tilde{B}$ \cite{Horn_85}, respectively, where $a$
and $b$ are local symplectic eigenvalues of CM $\gamma_{AB}$, Eq.~(\ref{gammast2}).
By exactly the same argument which leads to inequality
(\ref{nu12ineq}) we then have $a^2\geq\mbox{det}\tilde{A}$ and
$b^2\geq\mbox{det}\tilde{B}$, which finally implies the following
inequalities
\begin{equation}\label{tildeabineq}
a\geq\tilde{a},\quad b\geq\tilde{b},
\end{equation}
where we have used equalities
$\tilde{a}=\sqrt{\mbox{det}\tilde{A}}$ and
$\tilde{b}=\sqrt{\mbox{det}\tilde{B}}$.

If we now combine inequalities (\ref{nu12ineq}) and
(\ref{tildeabineq}) with inequality
\begin{equation}\label{arithgeom}
\sqrt{\frac{\tilde{a}}{\tilde{b}}}+\sqrt{\frac{\tilde{b}}{\tilde{a}}}=\frac{\tilde{a}+\tilde{b}}{\sqrt{\tilde{a}\tilde{b}}}\geq
2,
\end{equation}
which follows from the inequality of arithmetic and geometric
means, we arrive at a new lower bound $\tilde{G}_{\rm min}$ on
$\tilde{G}$, $\tilde{G}\geq\tilde{G}_{\rm min}$, of the form:
\begin{equation}\label{tildeGmin}
\tilde{G}_{\rm min}=2+\frac{1}{\sqrt{ab}}-\sqrt{\nu_{1}\nu_{2}},
\end{equation}
where $\nu_{1,2}$ are symplectic eigenvalues (\ref{nu12}) of the
investigated state $\rho_{AB}$. Hence, if for a two-mode Gaussian
state with standard-form CM (\ref{gammast2}) inequality
\begin{equation}\label{newGbound}
2+\frac{1}{\sqrt{ab}}\geq\sqrt{\nu_{1}\nu_{2}}
\end{equation}
is obeyed, $\tilde{G}\geq0$ for any CM $\Gamma_{E}$, and homodyne
detection of quadratures $x_{A}$ and $x_{B}$ on modes $A$ and $B$
is optimal for any $\Gamma_{E}$. The GCMI then always reads as
in Eq.~(\ref{IcGcondhom}), and for symmetric squeezed thermal states as
well as asymmetric squeezed thermal GLEMS the GIE is given by
Eqs.~(\ref{GIEsymTMST}) and (\ref{GIEasymGLEMS}), respectively.

Note first, that in contrast with the derivation of original
inequalities $a\leq 2.41$ and $\sqrt{ab}\leq2.41$, which utilized
a specific structure of states for which they were derived, no
similar restrictive assumptions have been made when deriving
inequality (\ref{newGbound}), and thus it holds for {\it any}
two-mode Gaussian state. Needles to say further, that inequality
$\tilde{G}_{\rm min}\geq0$ provides a strictly stronger condition,
i.e., it is satisfied by a strictly larger set of states, than
original inequalities. This is a consequence of inequality
\begin{equation}\label{nuab}
\sqrt{\nu_{1}\nu_{2}}=\sqrt[4]{\mbox{det}\gamma_{AB}}=\sqrt[4]{(ab-k_{x}^{2})(ab-k_{p}^{2})}<\sqrt{ab},
\end{equation}
where the strict inequality follows from inequality $k_{x}\geq
k_{p}>0$ given below Eq.~(\ref{gammast2}). Now, if we combine
Eq.~(\ref{tildeGmin}) with inequality (\ref{nuab}) we get
\begin{equation}\label{tildeGminstrict}
\tilde{G}_{\rm
min}>2\left(1-\frac{\sqrt{ab}-\frac{1}{\sqrt{ab}}}{2}\right).
\end{equation}
If now $\sqrt{ab}\leq2.41$, or $a\leq2.41$ for the case $a=b$, the
RHS of inequality (\ref{tildeGminstrict}) is
nonnegative, which implies $\tilde{G}_{\rm min}>0$, and thus
condition (\ref{newGbound}) is satisfied for all states for which the
original inequalities hold. Consider now an entangled symmetric
two-mode squeezed thermal state with parameters
$a=\sqrt{6}\doteq2.45$ and $k=2$. For this state inequality
$a\leq2.41$ is clearly not satisfied, whereas $\tilde{G}_{\rm
min}\doteq0.99>0$ still holds, and thus condition
(\ref{newGbound}) is indeed stronger than the original one.
Finally, the bound (\ref{tildeGmin}) is tight for some classes of
states but it is not tight always. For instance, for the class of
symmetric two-mode squeezed thermal states $\rho_{AB}^{(2)}$ the bound boils down to
\begin{equation}\label{tildeGmin2}
\tilde{G}_{\rm min}^{(2)}=2+\frac{1}{a}-\nu,
\end{equation}
which is tight, because it is reached by dropping the purifying
subsystem $E$, or equivalently, by projecting the subsystem onto
a product of two infinitely hot thermal states with CM $\Gamma_{E}^{\langle
n\rangle\rightarrow+\infty}$, where $\Gamma_{E}^{\langle
n\rangle}=(2\langle n\rangle+1)\openone_{4}$. On the other hand, in
the case of symmetric GLEMS one can minimize analytically
$\tilde{G}$, Eq.~(\ref{tildeG}), over all CMs $\Gamma_{E}$, which
yields another lower bound ($\equiv\tilde{G}_{\rm opt}$) of the
form \cite{Mista_16}
\begin{equation}\label{tildeGpom}
\tilde{G}_{\rm opt}\equiv2+\frac{1}{a}-\sqrt{a^{2}-k_{x}^{2}}.
\end{equation}
Consider now a mixed symmetric GLEMS, which has to fulfill
inequality $k_{x}>k_{p}$, because equality $k_{x}=k_{p}$ implies
purity of the state. Then, according to the latter inequality and
the left-hand side of inequality (\ref{nuab}), we have
\begin{equation}\label{nottight}
\sqrt{\nu_{1}}=\sqrt{\nu_{1}\nu_{2}}=\sqrt[4]{(a^2-k_{x}^{2})(a^2-k_{p}^{2})}>\sqrt{a^{2}-k_{x}^{2}},
\end{equation}
where equality $\nu_{2}=1$ was used. Hence, $\tilde{G}_{\rm
min}<\tilde{G}_{\rm opt}$, and the lower bound (\ref{tildeGmin})
is not tight.

In this section we have derived a sufficient condition for the
GCMI (\ref{IcG}) for a generic two-mode Gaussian state with CM
(\ref{gammast2}) to be always reached by double homodyne
detection. The condition attains a particularly simple form for
symmetric two-mode squeezed thermal states, when it simplifies to
\begin{equation}\label{tildeGmin2final}
\nu\leq 2+\frac{1}{a}.
\end{equation}
Because the condition is stronger than the original condition
$a\leq2.41$, our finding extends the formula for GIE,
Eq.~(\ref{GIEsymTMST}), to all symmetric two-mode squeezed thermal
states satisfying inequality (\ref{tildeGmin2final}).

A distinctive feature of condition (\ref{newGbound}) is that it is
valid for any two-mode Gaussian state. This gives us a prospect
that we will be able to calculate GIE even for more generic
two-mode Gaussian states, including those states with $a\ne b$ and
simultaneously $k_{x}\ne k_{p}$. To achieve this goal, we have to be able
to perform minimization on the RHS of Eq.~(\ref{Ubound}). By rewriting Eq.~(\ref{Ubound}) as
\begin{equation}\label{Uboundhom}
U\left(\rho_{AB}\right)=-\ln\sqrt{1-h_{\rm min}},
\end{equation}
where
\begin{equation}\label{hmin}
h_{\rm min}\equiv\mathop{\mbox{inf}}_{\Gamma_{E}}\left(\frac{\tilde{c}_{x}^2}{\tilde{a}\tilde{b}}\right),
\end{equation}
we see, that minimization in Eq.~(\ref{Ubound}) is equivalent with minimization on the RHS of Eq.~(\ref{hmin}).
Our ability to carry out the minimization on the RHS of the last formula strongly depends on the structure
of the investigated state $\rho_{AB}$. Previously \cite{Mista_15,Mista_16}, this approach proved to be successful in derivation of
the upper bound (\ref{Ubound}) for symmetric GLEMS and asymmetric two-mode squeezed thermal GLEMS.
Later in this paper we show, that the same method can be also used for evaluation of the upper bound on
GIE for several new classes of two-mode GLEMS. Before doing that, however, we first briefly explain the last step of
the method of calculation of GIE, that is, the saturation of the upper bound.

\section{SATURATION OF THE UPPER BOUND}\label{sec_saturation}

We now move to the description of a method, which allows us to show for
all states investigated here as well as in Ref.~\cite{Mista_16}, that the conditional mutual information
(\ref{IABE}) for homodyne detection of quadratures $x_{A}$ and $x_{B}$
on modes $A$ and $B$, respectively, which is minimized with respect
to all CMs $\Gamma_{E}$, saturates the upper bound (\ref{Ubound}).

For this purpose, it is convenient to express blocks $\gamma_{AB}$ and $\gamma_{ABE}$ of CM $\gamma_{\pi}$,
Eq.~(\ref{gammapi}), as
\begin{eqnarray}\label{gammaABgammaAEgammaBE}
\gamma_{AB}=\left(\begin{array}{cc}
\gamma_{A} & \omega_{AB}  \\
\omega_{AB}^{T} & \gamma_{B} \\
\end{array}\right),\quad \gamma_{ABE}=\left(\begin{array}{c}
\gamma_{AE}\\
\gamma_{BE}\\
\end{array}\right).
\end{eqnarray}
Next, we apply to the matrix
$\Sigma_{ABE}\equiv\gamma_{\pi}+\Gamma_{A}\oplus\Gamma_{B}\oplus\Gamma_{E}$
the determinant formula \cite{Horn_85}:
\begin{equation}\label{det}
\mbox{det}(M)=\mbox{det}(\mathfrak{D})\mbox{det}(\mathfrak{A}-\mathfrak{B}\mathfrak{D}^{-1}\mathfrak{C}),
\end{equation}
which is valid for any $(n+m)\times(n+m)$ matrix
\begin{eqnarray}\label{M}
M=\left(\begin{array}{cc}
\mathfrak{A} & \mathfrak{B}\\
\mathfrak{C} & \mathfrak{D}\\
\end{array}\right),
\end{eqnarray}
where $\mathfrak{A}$, $\mathfrak{B}$ and $\mathfrak{C}$ are
respectively $n\times n$, $n\times m$ and $m\times n$ matrices and
$\mathfrak{D}$ is an $m\times m$ invertible matrix. This allows us
to express the determinant of the correlation matrix (\ref{sigma}) as
\begin{equation}\label{detsigmaAB}
\mbox{det}\sigma_{AB}=\frac{\mbox{det}(\Gamma_{A}\oplus\Gamma_{B}+\gamma_{AB})\mbox{det}(\Gamma_{E}+X_{AB})}{\mbox{det}(\Gamma_{E}+\gamma_{E})},
\end{equation}
where
\begin{eqnarray}\label{XAB}
X_{AB}=\gamma_{E}-\gamma_{ABE}^{T}(\Gamma_{A}\oplus\Gamma_{B}+\gamma_{AB})^{-1}
\gamma_{ABE}.
\end{eqnarray}
Likewise, application of the determinant formula (\ref{det}) to the reduced
matrices $\Sigma_{jE}$, $j=A,B$, of subsystem $(jE)$, yields
\begin{equation}\label{detsigmaj}
\mbox{det}\sigma_{j}=\frac{\mbox{det}(\Gamma_{j}+\gamma_{j})\mbox{det}(\Gamma_{E}+X_{j})}{\mbox{det}(\Gamma_{E}+\gamma_{E})},
\end{equation}
where
\begin{eqnarray}\label{Xj}
X_{j}=\gamma_{E}-\gamma_{jE}^{T}(\Gamma_{j}+\gamma_{j})^{-1}\gamma_{jE}.
\end{eqnarray}
Hence, the conditional mutual information (\ref{IABE}) can be rewritten into the form
\begin{eqnarray}\label{IABEgen}
I(A;B|E)=I(A;B)+K(E|A;B),
\end{eqnarray}
where $I(A;B)$ is given in Eq.~(\ref{IAB}) and
\begin{equation}\label{KEABhelp}
K(E|A;B)=\frac{1}{2}\ln\mathscr{K}
\end{equation}
with
\begin{eqnarray}\label{K1}
\mathscr{K}=\frac{\mbox{det}(\Gamma_{E}+X_{A})\mbox{det}(\Gamma_{E}+X_{B})}{\mbox{det}(\Gamma_{E}+X_{AB})\mbox{det}(\Gamma_{E}+\gamma_{E})}.
\end{eqnarray}

The expression of the conditional mutual information given on the
RHS of Eq.~(\ref{IABEgen}) simplifies its minimization over all CMs $\Gamma_{E}$.
Consider now homodyne detection of quadratures $x_{A}$ and $x_{B}$ on modes $A$ and $B$,
which is described by CMs $\Gamma_{A}^{t}\equiv\mbox{diag}(e^{-2t},e^{2t})$
and $\Gamma_{B}^{t}\equiv\mbox{diag}(e^{-2t},e^{2t})$
in the limit $t\rightarrow+\infty$. In this case Eq.~(\ref{IABEgen}) boils down to
\begin{eqnarray}\label{IcGcondhom2}
I_{h}\left(A;B|E\right)=\mathcal{I}_{c,h}^{G}\left(\rho_{AB}\right)+K_{h}(E|A;B),
\end{eqnarray}
where $\mathcal{I}_{c,h}^{G}\left(\rho_{AB}\right)$ is obtained from Eq.~(\ref{IcGhom}) by replacing $c_{x}$ with $k_{x}$, and
\begin{equation}\label{KEABhom}
K_{h}(E|A;B)=\frac{1}{2}\ln \mathscr{K}_{h},
\end{equation}
where $\mathscr{K}_{h}$ is obtained from the RHS of
Eq.~(\ref{K1}), by putting $\Gamma_{A}=\Gamma_{A}^{t}$ and
$\Gamma_{B}=\Gamma_{B}^{t}$ and taking the limit
$t\rightarrow+\infty$.

The remaining step is the minimization of the conditional mutual information (\ref{IcGcondhom2}) over all CMs $\Gamma_{E}$, which
boils down to finding of the quantity
\begin{eqnarray}\label{L}
L(\rho_{AB})&\equiv&\mathop{\mbox{inf}}_{\Gamma_{E}}\left[I_{h}\left(A;B|E\right)\right]\nonumber\\
&=&\frac{1}{2}\ln\left(\frac{ab}{ab-k_{x}^{2}}\right)+\frac{1}{2}\ln\mathscr{K}_{\rm
min},
\end{eqnarray}
where
\begin{equation}\label{Kmin}
\mathscr{K}_{\rm
min}\equiv\mathop{\mbox{inf}}_{\Gamma_{E}}\mathscr{K}_{h},
\end{equation}
and where we used Eq.~(\ref{IcGhom}) and monotonicity of the logarithmic
function. Now, if for some state $\rho_{AB}$ the quantity (\ref{L}) is equal to the upper bound (\ref{Ubound}),
we found for fixed measurements on modes $A$ and $B$ the minimal conditional mutual information (\ref{IABE}) with
respect to all CMs $\Gamma_{E}$, which cannot be improved, and thus the upper bound coincides with GIE.
In what follows we illustrate the utility of this approach for derivation of GIE for several new
classes of GLEMS.

\section{GIE FOR GLEMS}\label{sec_GIE_for_GLEMS}

As we have already mentioned, GLEMS are Gaussian states with minimal negativity for fixed global and local purities
\cite{Adesso_04,Adesso_04b}, and they naturally appear in a cryptographic setting involving two-mode
squeezed vacuum (\ref{TMSV}) with one mode transmitted through a purely lossy channel.
Here, we restrict ourself to a subset of GLEMS with CM (\ref{gammast2}), which is characterized by
condition $a+b-1>\sqrt{\mbox{det}\gamma_{AB}}$ \cite{Adesso_04}, and which is relevant for calculation of GIE
because it contains all entangled GLEMS. The GLEMS from the subset, which we call from now simply as GLEMS for brevity,
saturate the first of inequalities (\ref{CMcondition}) which express the Heisenberg
uncertainty principle, and thus they are states with partial minimum uncertainty. From Eqs.~(\ref{nu12})
and (\ref{DeltaD}) it follows, that saturation of the first of inequalities (\ref{CMcondition}) is equivalent
with equality $\nu_{2}=1$, whereas the other symplectic eigenvalue is equal to
\begin{equation}\label{nu}
\nu\equiv\nu_{1}=\sqrt{\mbox{det}\gamma_{AB}}.
\end{equation}
The symplectic matrix which brings CM (\ref{gammast2}) for GLEMS to Williamson normal form (\ref{Williamson}) reads as
\begin{eqnarray}\label{Sgeneric2}
S=\left(\begin{array}{cccc}
x_1 & 0 & x_2 & 0\\
0 & x_3 & 0 & x_4 \\
x_5 & 0 & x_6 & 0 \\
0 & x_7 & 0 & x_8 \\
\end{array}\right),
\end{eqnarray}
where the explicit expression of matrix elements $x_{1}, x_{2},\ldots, x_{8}$ in
terms of parameters $a, b, k_{x}$ and $k_{p}$ is given in the Appendix~\ref{Sdiagonalization}.

Since we already know GIE for GLEMS with $\nu=1$, which coincide with pure states, symmetric
GLEMS with $a=b$ and $\nu>1$, as well as asymmetric GLEMS with $a\ne b$ and $k_{x}=k_{p}$,
here we focus on derivation of GIE for several other classes of GLEMS with $\nu>1$. Derivation of
the symplectic matrix which brings CM (\ref{gammast2})
to the Williamson normal form performed in Appendix~\ref{Sdiagonalization}
unveils, that depending on the relation among parameters $a,b,k_{x}$ and $k_{p}$ we have
to distinguish another four sets of GLEMS. This includes GLEMS ($\equiv\rho_{AB}^{(4)}$)
with $a>b$ and $bk_{x}=ak_{p}$, ($\equiv\rho_{AB}^{(5)}$) with $a<b$ and $ak_{x}=bk_{p}$,
($\equiv\rho_{AB}^{(6)}$) with $a>b$ and $bk_{x}\ne ak_{p}$, and ($\equiv\rho_{AB}^{(7)}$)
with $a<b$ and $ak_{x}\ne bk_{p}$. In what follows, we compute an analytical formula for GIE of all
states $\rho_{AB}^{(4)}$ and $\rho_{AB}^{(5)}$ satisfying condition (\ref{newGbound}). Moreover,
we also outline how to calculate GIE for states $\rho_{AB}^{(6)}$ and $\rho_{AB}^{(7)}$
obeying inequality (\ref{newGbound}), by calculating it explicitly for a particular example of a state
$\rho_{AB}^{(6)}$.

\subsection{Upper bound for GLEMS}\label{sec_U_GLEMS}

For evaluation of the upper bound on GIE, Eq.~(\ref{Ubound}), we need to calculate the quantity (\ref{hmin}),
which requires to express parameters of the standard-form CM (\ref{gammacondst}) as functions of parameters of CM $\Gamma_{E}$.
Owing to condition $\nu_{2}=1$, GLEMS possess unit symplectic rank, $R=1$, and thus their Williamson normal
form (\ref{Williamson}) reads as $\gamma_{AB}^{(0)}=(\nu\openone)\oplus\openone$. In addition, from equation
(\ref{gammaABEgammaE0}) it follows that
\begin{eqnarray}\label{GLEMSgammaABEgammaE0}
\gamma_{ABE}^{(0)}=\left(\begin{array}{c}
\sqrt{\nu^{2}-1}\sigma_{z} \\
\mathbb{O}_{2\times2} \\
\end{array}\right),\quad\gamma_{E}^{(0)}=\nu\openone,
\end{eqnarray}
which reveals that the purifying subsystem $E$ is single-mode.
This allows us to take CM $\Gamma_{E}$ appearing in Eq.~(\ref{gammacond0}) in the form:
\begin{equation}\label{GammaE}
\Gamma_E=P(\varphi)\mbox{diag}(V_{x},V_{p})P^{T}(\varphi),
\end{equation}
where
\begin{equation}\label{P}
P(\varphi)= \left(
\begin{array}{cc}
\cos\varphi  & -\sin \varphi \\
\sin\varphi & \cos\varphi \\
\end{array}
\right)
\end{equation}
with $\varphi\in[0,\pi)$, $V_{x}=\tau e^{2t}$ and $V_{p}=\tau e^{-2t}$, where $\tau\geq1$ and $t\geq0$.
By calculating the inverse matrix on the RHS of Eq.~(\ref{gammacond0}), and making use of Eqs.~(\ref{GLEMSgammaABEgammaE0})
together with relation $P^{T}(\varphi)=\sigma_{z}P(\varphi)\sigma_{z}$, we get after some algebra CM (\ref{gammacondS})
in the form  \cite{Mista_16}:
\begin{eqnarray}\label{gammacondS2}
\gamma_{AB|E}=S^{-1}\left(\gamma_{A|E}\oplus\openone_{B}\right)(S^{-1})^{T}.
\end{eqnarray}
Here,
\begin{eqnarray}\label{gammaAE}
\gamma_{A|E}&=&P^{T}(\varphi)\mbox{diag}(\mathcal{V}_{x},\mathcal{V}_{p})P(\varphi)\nonumber\\
&=&\left(\begin{array}{cc}
\mathcal{V}_{+}+\mathcal{V}_{-}\cos(2\varphi)  & -\mathcal{V}_{-}\sin(2\varphi) \\
-\mathcal{V}_{-}\sin(2\varphi) & \mathcal{V}_{+}-\mathcal{V}_{-}\cos(2\varphi) \\
\end{array}
\right),\nonumber\\
\end{eqnarray}
with $\mathcal{V}_{\pm}=(\mathcal{V}_{x}\pm\mathcal{V}_{p})/2$, where
\begin{equation}\label{calVxVp}
\mathcal{V}_{x}=\frac{\nu V_{x}+1}{\nu+V_{x}},\quad
\mathcal{V}_{p}=\frac{\nu V_{p}+1}{\nu+V_{p}},
\end{equation}
$\nu\geq\mathcal{V}_{x}\geq\mathcal{V}_{p}\geq1/\nu$, are eigenvalues of CM (\ref{gammaAE}), and
\begin{eqnarray}\label{invSgeneric}
S^{-1}=\left(\begin{array}{cccc}
x_3 & 0 & x_7 & 0\\
0 & x_1 & 0 & x_5 \\
x_4 & 0 & x_8 & 0 \\
0 & x_2 & 0 & x_6 \\
\end{array}\right)
\end{eqnarray}
is the inverse of symplectic matrix (\ref{Sgeneric2}), which can be calculated with the help of formula $S^{-1}=\Omega_{2} S^{T}\Omega_{2}^{T}$.
If we now substitute matrix (\ref{invSgeneric}) into the RHS of Eq.~(\ref{gammacondS2}) and we express the obtained matrix
in the block form (\ref{gammacondblock}), we can calculate all parameters needed for calculation of the quantity $h_{\rm min}$, Eq.~(\ref{hmin}).
Below we will see, that all we need are parameters $\tilde{a}^{2}, \tilde{b}^{2}$ and $\tilde{c}_{x}\tilde{c}_{p}$, which can be obtained
from the formulas $\tilde{a}^{2}=\mbox{det}\tilde{A}, \tilde{b}^{2}=\mbox{det}\tilde{B}$
and $\tilde{c}_{x}\tilde{c}_{p}=\mbox{det}\tilde{C}$ in the form
\begin{eqnarray}\label{tildeabcxcp}
\tilde{a}^{2}&=&x_{1}^{2}x_{3}^{2}\mathcal{V}_{x}\mathcal{V}_{p}+x_{3}^{2}x_{5}^{2}\gamma_{11}+x_{1}^{2}x_{7}^{2}\gamma_{22}+x_{5}^{2}x_{7}^{2},\nonumber\\
\tilde{b}^{2}&=&x_{2}^{2}x_{4}^{2}\mathcal{V}_{x}\mathcal{V}_{p}+x_{4}^{2}x_{6}^{2}\gamma_{11}+x_{2}^{2}x_{8}^{2}\gamma_{22}+x_{6}^{2}x_{8}^{2},\nonumber\\
\tilde{c}_{x}\tilde{c}_{p}&=&x_{1}x_{2}x_{3}x_{4}\mathcal{V}_{x}\mathcal{V}_{p}+x_{3}x_{4}x_{5}x_{6}\gamma_{11}\nonumber\\
&&+x_{1}x_{2}x_{7}x_{8}\gamma_{22}+x_{5}x_{6}x_{7}x_{8},
\end{eqnarray}
where we have set $\gamma_{ii}\equiv(\gamma_{A|E})_{ii}$, $i=1,2$, for the sake of simplicity. To proceed further
with calculation of the upper bound (\ref{Ubound}) we have to express parameters $x_{1},x_{2},\ldots,x_{8}$ appearing
in Eqs.~(\ref{tildeabcxcp}) via parameters $a,b,k_{x}$ and $k_{p}$. This requires to distinguish the following cases:

\subsubsection{GLEMS with $a>b$ and $bk_{x}=ak_{p}$}

Let us consider first GLEMS $\rho_{AB}^{(4)}$ with $a>b$ and $bk_{x}=ak_{p}$. This class of states
is relevant from the point of view of calculation of the upper bound on GIE based on formula (\ref{IcGcondhom}),
because there exist physical entangled  GLEMS satisfying equation $bk_{x}=ak_{p}$, for which inequality
(\ref{newGbound}) is obeyed. Indeed, consider a matrix (\ref{gammast2})
with $a=2\sqrt{2}, b=k_{x}=\sqrt{2}$ and $k_{p}=1/\sqrt{2}$, which clearly satisfies equality $bk_{x}=ak_{p}$.
The matrix also describes a physical entangled state, because both state conditions (\ref{CMcondition}) as well as
entanglement condition (\ref{entcondition}) are fulfilled. Additionally, one has $\nu_{2}=\sqrt{b^2-k_{x}k_{p}}=1$,
Eq.~(\ref{nuAnuB2a}) of Appendix~\ref{Sdiagonalization}, and thus the state is GLEMS. Finally, for the lower bound (\ref{tildeGmin}) one
gets $\tilde{G}_{\rm min}=2.5-\sqrt[4]{7}\doteq 0.873>0$, which implies that double homodyning on modes $A$ and $B$ is
optimal for the present state, and therefore the upper bound (\ref{Ubound}) can be calculated by carrying out minimization on
the RHS of Eq.~(\ref{hmin}).

For this purpose, we calculate parameters (\ref{tildeabcxcp}) which attain a particularly simple form. Indeed, from Eq.~(\ref{S2a})
of Appendix~\ref{Sdiagonalization} one finds that
\begin{eqnarray}\label{xclass1}
x_{1}&=&\sqrt{\frac{\nu}{a}},\quad x_{2}=0,\quad x_{3}=\sqrt{\frac{a}{\nu}},\quad x_{4}=\frac{k_{x}}{\sqrt{a\nu}},\nonumber\\
x_{5}&=&-\frac{k_{p}}{\sqrt{b}},\quad x_{6}=\sqrt{b},\quad x_{7}=0,\quad x_{8}=\frac{1}{\sqrt{b}},
\end{eqnarray}
where
\begin{equation}\label{nu4}
\nu=\sqrt{a^2-k_{x}k_{p}},
\end{equation}
and hence Eqs.~(\ref{tildeabcxcp}) yield
\begin{eqnarray}\label{tildeabcxcp1}
\tilde{a}^{2}&=&\mathcal{V}_{x}\mathcal{V}_{p}+\frac{k_{x}k_{p}}{\nu}[\mathcal{V}_{+}+\mathcal{V}_{-}\cos(2\varphi)],\nonumber\\
\tilde{b}^{2}&=&1+\frac{k_{x}k_{p}}{\nu}[\mathcal{V}_{+}+\mathcal{V}_{-}\cos(2\varphi)],\nonumber\\
\tilde{c}_{x}\tilde{c}_{p}&=&-\frac{k_{x}k_{p}}{\nu}[\mathcal{V}_{+}+\mathcal{V}_{-}\cos(2\varphi)],
\end{eqnarray}
where we used matrix (\ref{gammaAE}) and condition $bk_{x}=ak_{p}$.

Moving to minimization on the RHS of Eq.~(\ref{hmin}) one can see, that it can be carried out using the following chain of inequalities:
\begin{eqnarray}\label{h1}
\frac{\tilde{c}_{x}^2}{\tilde{a}\tilde{b}}&\geq&-\frac{\tilde{c}_{x}\tilde{c}_{p}}{\tilde{a}\tilde{b}}\geq -\frac{\tilde{c}_{x}\tilde{c}_{p}}{\tilde{a}^{2}}\nonumber\\
&=&1-\frac{1}{1+\frac{k_{x}k_{p}}{\nu\mathcal{V}_{x}\mathcal{V}_{p}}[\mathcal{V}_{+}+\mathcal{V}_{-}\cos(2\varphi)]}\nonumber\\
&\geq&1-\frac{1}{1+\frac{k_{x}k_{p}}{\nu\mathcal{V}_{x}}}\geq1-\frac{1}{1+\frac{k_{x}k_{p}}{\nu^{2}}}=\frac{k_{x}k_{p}}{a^{2}}.
\end{eqnarray}
Here, the first inequality is a consequence of inequality $\tilde{c}_{x}\tilde{c}_{p}<0$, which follows from the third of
Eqs.~(\ref{tildeabcxcp1}), and inequality $\tilde{c}_x\geq|\tilde{c}_{p}|\geq0$ given below Eq.~(\ref{gammacondst}), whereas
the second inequality steams from inequality $\tilde{a}\geq\tilde{b}$ resulting from the first two of Eqs.~(\ref{tildeabcxcp1}) and the fact that $\mathcal{V}_{x}\mathcal{V}_{p}\geq 1$. Further, the third inequality is
obtained if we notice that since $\mathcal{V}_{x}\geq\mathcal{V}_{p}$ one has $\mathcal{V}_{-}\geq0$, and thus the
expression in the square brackets is minimized for $\varphi=\pi/2$. Finally, as $\mathcal{V}_{x}\leq\nu$ the last
inequality is satisfied, while Eq.~(\ref{nu4}) has been used to obtain the last equality.

Importantly, the lower bound (\ref{h1}) is tight, because it is reached for CM (\ref{GammaE}), e.g., with $\varphi=\pi/2,
V_{x}V_{p}=1$ and in the limit $t\rightarrow+\infty$, which corresponds to homodyne detection of quadrature
$x_{E}$ on mode $E$. As a result, one gets $h_{\rm min}=k_{x}k_{p}/a^{2}$, Eq.~(\ref{hmin}), which gives after the substitution into the
RHS of Eq.~(\ref{Uboundhom}) the upper bound we are looking for,
\begin{equation}\label{U4}
U\left(\rho_{AB}^{(4)}\right)=\ln\left(\frac{a}{\nu}\right).
\end{equation}

\subsubsection{GLEMS with $a<b$ and $ak_{x}=bk_{p}$}

Derivation of the upper bound (\ref{Ubound}) for GLEMS $\rho_{AB}^{(5)}$ with $a<b$ and $ak_{x}=bk_{p}$ closely follows derivation
performed in previous case. First, it is straightforward to find an example of a state from the considered class of states.
Namely, owing to symmetry of conditions (\ref{CMcondition}) and (\ref{entcondition}) with respect
to exchange $a\leftrightarrow b$ it is obvious, that both the conditions are satisfied also by a CM obtained
from CM of previous example by exchanging the values of parameters $a$ and $b$, i.e., by a CM (\ref{gammast2})
with $b=2\sqrt{2}$, $a=k_{x}=\sqrt{2}$ and $k_{p}=1/\sqrt{2}$. It is also clear that the new CM fulfils both other
conditions $a<b$ and $ak_{x}=bk_{p}$. Moreover, since states satisfying the latter two conditions possess symplectic
eigenvalues
\begin{equation}\label{tildenu5}
\tilde{\nu}\equiv\nu_{1}=\sqrt{b^2-k_{x}k_{p}}
\end{equation}
and $\nu_{2}=\sqrt{a^2-k_{x}k_{p}}$, Eq.~(\ref{nuAnuB3a}) of Appendix~\ref{Sdiagonalization}, the new CM has
the same symplectic eigenvalues as the original CM and thus it describes a GLEMS with a strictly positive lower bound
(\ref{tildeGmin}). As a result, the new CM is again the sought example of a state from
the investigated class of states, for which the upper bound (\ref{Ubound}) can be obtained by calculating
the quantity (\ref{hmin}).

Further, making use of equation (\ref{S3a}) of Appendix~\ref{Sdiagonalization} one finds the elements of
symplectic matrix (\ref{invSgeneric}) to be
\begin{eqnarray}\label{xclass2}
x_{1}&=&0,\quad x_{2}=\sqrt{\frac{\tilde{\nu}}{b}},\quad x_{3}=\frac{k_{x}}{\sqrt{b\tilde{\nu}}},\quad x_{4}=\sqrt{\frac{b}{\tilde{\nu}}},\nonumber\\
x_{5}&=&\sqrt{a},\quad x_{6}=-\frac{k_{p}}{\sqrt{a}},\quad x_{7}=\frac{1}{\sqrt{a}},\quad x_{8}=0,\nonumber\\
\end{eqnarray}
which gives after substitution into RHS of Eq.~(\ref{tildeabcxcp})
\begin{eqnarray}\label{tildeabcxcp2}
\tilde{a}^{2}&=&1+\frac{k_{x}k_{p}}{\tilde{\nu}}[\mathcal{V}_{+}+\mathcal{V}_{-}\cos(2\varphi)],\nonumber\\
\tilde{b}^{2}&=&\mathcal{V}_{x}\mathcal{V}_{p}+\frac{k_{x}k_{p}}{\tilde{\nu}}[\mathcal{V}_{+}+\mathcal{V}_{-}\cos(2\varphi)],\nonumber\\
\tilde{c}_{x}\tilde{c}_{p}&=&-\frac{k_{x}k_{p}}{\tilde{\nu}}[\mathcal{V}_{+}+\mathcal{V}_{-}\cos(2\varphi)].
\end{eqnarray}
Comparison of the latter parameters with parameters (\ref{tildeabcxcp1}) unveils that the former can
be obtained from the latter by replacing $\nu$ with $\tilde{\nu}$ and exchanging the right-hand sides of
equations for $\tilde{a}^{2}$ and $\tilde{b}^{2}$. Repeating the same procedure as that of leading to
the lower bound (\ref{h1}) we get the same chain of inequalities as in (\ref{h1}) just with $\tilde{a}^{2}$ replaced
with $\tilde{b}^{2}$ on the RHS of the second inequality, $\nu$ replaced with $\tilde{\nu}$ in the remaining inequalities,
and $a$ replaced with $b$ in the final lower bound. Thus one finds the lower bound in the following form:
\begin{eqnarray}\label{h2}
\frac{\tilde{c}_{x}^2}{\tilde{a}\tilde{b}}\geq\frac{k_{x}k_{p}}{b^{2}},
\end{eqnarray}
which is again saturated by homodyne detection of quadrature $x_{E}$.
Hence, from Eq.~(\ref{Uboundhom}) we immediately arrive at the upper bound
\begin{equation}\label{U5}
U\left(\rho_{AB}^{(5)}\right)=\ln\left(\frac{b}{\tilde{\nu}}\right).
\end{equation}

\subsubsection{Generic GLEMS}\label{subsubsec_U_generic_GLEMS}

Previous method of derivation of the upper bound (\ref{Ubound}) can be extended
to more generic GLEMS the parameters of which do not satisfy any additional condition except for
the defining equality $\nu_{2}=1$. To illustrate this, we calculate
the bound for one example of a state $\rho_{AB}^{(6)}$, which also
gives us a recipe of how to evaluate the bound for some other states
$\rho_{AB}^{(6)}$ and $\rho_{AB}^{(7)}$.

The example state ($\equiv\tilde{\rho}_{AB}^{(6)}$) has a CM (\ref{gammast2}) with parameters
$a=2\sqrt{2}$, $b=\sqrt{2}$ and $k_{x,p}=(\sqrt{97}\pm1)/8$. As the first
of inequalities (\ref{CMcondition}) is saturated whereas the second one is
fulfilled owing to inequality $ab-k_{x}^2=(79-\sqrt{97})/32\doteq 2.161>1$, the considered
CM describes a physical GLEMS. Further, because $a>b$ and
$\tilde{M}=bk_{x}-ak_{p}=(3-\sqrt{97})/(4\sqrt{2})\doteq -1.211<0$, the state belongs to the
class of states $\rho_{AB}^{(6)}$. Finally, since inequality (\ref{entcondition})
boils down to inequality $7<13$, the GLEMS is entangled, and as $\tilde{G}_{\rm min}=5/2-\sqrt[4]{6}\doteq0.935>0$,
Eq.~(\ref{tildeGmin}), the upper bound (\ref{Ubound}) can be calculated using formulas
(\ref{Uboundhom}) and (\ref{hmin}).

To calculate the bound we can proceed analogously as in previous two cases.
First, we use conditions
\begin{eqnarray}\label{xconditions1}
x_{1}x_{3}+x_{2}x_{4}&=&1,\quad x_{1}x_{7}+x_{2}x_{8}=0,\nonumber\\
x_{5}x_{7}+x_{6}x_{8}&=&1,\quad x_{3}x_{5}+x_{4}x_{6}=0,
\end{eqnarray}
being a consequence of the symplectic condition (\ref{Scondition}) with $N=2$,
and conditions
\begin{equation}\label{xconditions2}
x_{1}x_{3}+x_{5}x_{7}=1,\quad x_{2}x_{4}+x_{6}x_{8}=1,
\end{equation}
which follow from equation $S^{-1}S=\openone_{4}$. Next, with the help of the conditions we can
express the product $\tilde{c}_{x}\tilde{c}_{p}$,
Eq.~(\ref{tildeabcxcp}), as
\begin{eqnarray}\label{tildecxcp6}
\tilde{c}_{x}\tilde{c}_{p}&=&x_{1}x_{3}x_{5}x_{7}(\mathcal{V}_{x}\mathcal{V}_{p}+1)-x_{3}^{2}x_{5}^{2}\gamma_{11}
-x_{1}^{2}x_{7}^{2}\gamma_{22}\nonumber\\
&=&x_{1}x_{3}\mathcal{V}_{x}\mathcal{V}_{p}+x_{5}x_{7}-\tilde{a}^{2},
\end{eqnarray}
where $\tilde{a}^{2}$ is given in Eq.~(\ref{tildeabcxcp}). From the explicit form of
parameters $x_{1}, x_{2},\ldots, x_{8}$ given in Eq.~(\ref{2b}) of the Appendix~\ref{Sdiagonalization}, one
can further find that
\begin{equation}\label{x1x3x5x7}
x_{1}x_{3}x_{5}x_{7}=\frac{M\tilde{M}}{D}<0,
\end{equation}
where quantities $D, M$ and $\tilde{M}$ are defined in
Eqs.~(\ref{DeltaD}) and (\ref{MtM}),
respectively. Consequently, first of Eqs.~(\ref{tildecxcp6}) reveals
that $\tilde{c}_{x}\tilde{c}_{p}<0$ and if we take into account condition $\tilde{c}_{x}\geq|\tilde{c}_{p}|\geq0$, equality
$\tilde{a}^{2}-\tilde{b}^{2}=(x_{1}x_{3}-x_{5}x_{7})(\mathcal{V}_{x}\mathcal{V}_{p}-1)$, and the second
equation in Eq.~(\ref{tildecxcp6}), we find after some algebra for the quantity $\tilde{c}_{x}^{2}/(\tilde{a}\tilde{b})$ to be
minimized the following lower bound:
\begin{eqnarray}\label{h3}
\frac{\tilde{c}_{x}^2}{\tilde{a}\tilde{b}}&\geq&-\frac{\tilde{c}_{x}\tilde{c}_{p}}{\tilde{a}\tilde{b}}=
\left\{\left[1+\frac{x_{1}x_{3}\mathcal{V}_{x}\mathcal{V}_{p}+x_{5}x_{7}}{\tilde{a}^{2}-(x_{1}x_{3}
\mathcal{V}_{x}\mathcal{V}_{p}+x_{5}x_{7})}\right]\right.\nonumber\\
&&\times\left.\left[1+\frac{x_{5}x_{7}\mathcal{V}_{x}\mathcal{V}_{p}+x_{1}x_{3}}{\tilde{a}^{2}-(x_{1}x_{3}
\mathcal{V}_{x}\mathcal{V}_{p}+x_{5}x_{7})}\right]\right\}^{-\frac{1}{2}}\equiv h. \nonumber\\
\end{eqnarray}
Our goal is now to minimize function $h$ over $\varphi\in[0,\pi)$ and eigenvalues $\mathcal{V}_{x},\mathcal{V}_{p}$, Eq.~(\ref{calVxVp}),
such that if $\mathcal{V}_{p}\in[1/\nu,1]$, then $\mathcal{V}_{x}\in[1/\mathcal{V}_{p},\nu]$, whereas if
$\mathcal{V}_{p}\in(1,\nu]$, then $\mathcal{V}_{x}\in[\mathcal{V}_{p},\nu]$, where $\nu$ is the larger symplectic eigenvalue (\ref{nu}) of
the considered state. It is more convenient to introduce new variables $\kappa\equiv\sqrt{\mathcal{V}_{x}\mathcal{V}_{p}}$ and $z\equiv\sqrt{\mathcal{V}_{x}/\mathcal{V}_{p}}$, which lie in the intervals $\kappa\in[1,\nu]$ and $z\in[1,\nu/\kappa]$ \cite{Mista_15},
and to carry out the minimization with respect to them. Now, making use of the first of conditions (\ref{xconditions2}),
relation $x_{1}x_{3}-x_{5}x_{7}=(a^{2}-b^{2})/\sqrt{D}$, and inequality $a>b$, one finds that $x_{1}x_{3}>0$
and hence utilizing Eq.~(\ref{x1x3x5x7}) it also follows that $x_{5}x_{7}<0$. This implies using inequality $\kappa\geq 1$ that
$x_{1}x_{3}\kappa^2+x_{5}x_{7}\geq x_{1}x_{3}+x_{5}x_{7}=1>0$. Likewise, as $\kappa\leq \nu$ one gets for the present
state $x_{5}x_{7}\kappa^{2}+x_{1}x_{3}\geq x_{5}x_{7}\nu^{2}+x_{1}x_{3}=1/2>0$ and both second terms in square brackets on the
RHS of Eq.~(\ref{h3}) are positive. Obviously, both the terms depend on variables $\varphi$ and $z$ only through the
parameter $\tilde{a}^{2}=\tilde{a}^{2}(\varphi,z,\kappa)$ and because they are both positive, minimization of function
$h=h(\varphi,z,\kappa)$, Eq.~(\ref{h3}), with respect to the variables can be performed by minimization of the parameter $\tilde{a}^{2}$.
By substituting for elements $\gamma_{11}$ and $\gamma_{22}$ from Eq.~(\ref{gammaAE}) into the RHS of expression for $\tilde{a}^{2}$,
Eq.~(\ref{tildeabcxcp}), we get
\begin{eqnarray}\label{tildea21}
\tilde{a}^{2}(\varphi,z,\kappa)&=&x_{1}^{2}x_{3}^{2}\kappa^{2}+x_{5}^{2}x_{7}^{2}+(x_{3}^{2}x_{5}^{2}+x_{1}^2x_{7}^{2})\mathcal{V}_{+}\nonumber\\
&&+(x_{3}^2x_{5}^{2}-x_{1}^{2}x_{7}^{2})\mathcal{V}_{-}\cos(2\varphi),
\end{eqnarray}
where parameters $\mathcal{V}_{\pm}$ are defined below Eq.~(\ref{gammaAE}). The minimization of $\tilde{a}^{2}$ with
respect to $\varphi$ is now straightforward. Namely, if we note that $x_{3}^2x_{5}^{2}-x_{1}^{2}x_{7}^{2}=ab(k_{x}^{2}-k_{p}^{2})/\sqrt{D\mbox{det}\gamma_{AB}}>0$ and $\mathcal{V}_{-}\geq0$, we immediately
see, that the minimum is reached for $\varphi=\pi/2$ and it reads as
\begin{eqnarray}\label{tildea2optphi}
\tilde{a}^{2}\left(\frac{\pi}{2},z,\kappa\right)&=&x_{1}^{2}x_{3}^{2}\kappa^{2}+x_{5}^{2}x_{7}^{2}
+\kappa\left(x_{1}^{2}x_{7}^{2}z+\frac{x_{3}^2x_{5}^{2}}{z}\right).\nonumber\\
\end{eqnarray}

It is also easy to minimize the latter quantity with respect to variable $z$. By solving
extremal equation $\partial\tilde{a}^{2}\left(\pi/2,z,\kappa\right)/\partial z=0$, one finds two stationary points
$z_{1,2}=\pm|x_{3}x_{5}/(x_{1}x_{7})|=\pm(x_{5}^2/x_{1}^{2})|x_{1}x_{3}/(x_{5}x_{7})|=\mp x_{3}x_{5}/(x_{1}x_{7})\doteq\pm2.269$.
Firstly, because $z_{2}<0$ one has $z_{2}\notin[1,\nu/\kappa]$.
Further, the other stationary point satisfies $z_{1}<\nu=\sqrt{6}\doteq 2.45$ and therefore $z_{1}\in[1,\nu/\kappa]$ for $\kappa\in[1,\nu/z_{1}]$,
while $z_{1}\notin[1,\nu/\kappa]$ for $\kappa\in(\nu/z_{1},\nu]$. Moreover, since for $z<z_{1}$ quantity
(\ref{tildea2optphi}) is a monotonically decreasing function of $z$
whereas for $z>z_{1}$ it is a monotonically increasing function, for $\kappa\in[1,\nu/z_{1}]$ it attains minimum at
$z_{1}$ of
\begin{equation}\label{tildea2optphiz1}
\tilde{a}^{2}\left(\frac{\pi}{2},z_{1},\kappa\right)=(x_{1}x_{3}\kappa-x_{5}x_{7})^{2}.
\end{equation}
On the other hand, for $\kappa\in(\nu/z_{1},\nu]$ the quantity $\tilde{a}^{2}$ reaches the minimal value on the boundary curve
$z=\nu/\kappa$ and it is given by
\begin{eqnarray}\label{tildea2optphiz2}
\tilde{a}^{2}\left(\frac{\pi}{2},\frac{\nu}{\kappa},\kappa\right)&=&\tilde{a}^{2}\left(\frac{\pi}{2},z_{1},\kappa\right)
+\left(\frac{x_{3}x_{5}}{\sqrt{\nu}}\kappa+x_{1}x_{7}\sqrt{\nu}\right)^{2}.\nonumber\\
\end{eqnarray}

In the final step of derivation of a tight lower bound on the quantity $\tilde{c}_{x}^{2}/(\tilde{a}\tilde{b})$ we
need to minimize function $h(\pi/2,z_{1},\kappa)$, where $h$ is defined in Eq.~(\ref{h3}), with respect to $\kappa\in[1,\nu/z_{1}]$,
and function $h(\pi/2,\nu/\kappa,\kappa)$ with respect to $\kappa\in(\nu/z_{1},\nu]$. In the first
case a rather lengthy algebra unveils that
\begin{equation}\label{h1x}
h\left(\frac{\pi}{2},z_{1},\kappa\right)=\left[1-\frac{\kappa}{x_{1}x_{3}x_{5}x_{7}(\kappa+1)^{2}}\right]^{-1}.
\end{equation}
Due to the fact that for $\kappa\geq1$ the function $\kappa/(\kappa+1)^{2}$ is monotonically decreasing and the expression $x_{1}x_{3}x_{5}x_{7}$
is negative, function (\ref{h1x}) is minimized at the boundary point $\kappa=1$, where it is equal to
\begin{equation}\label{hmin1}
h_{\rm min}^{(1)}\equiv h\left(\frac{\pi}{2},z_{1},1\right)=-\frac{4M\tilde{M}}{(a^2-b^2)^{2}}=\frac{11}{36}\doteq0.3056,
\end{equation}
where we used Eq.~(\ref{DeltaD}).

Moving to the minimization of function $h(\pi/2,\nu/\kappa,\kappa)$, it is clear from Eq.~(\ref{tildea2optphiz2}),
that on the interval $\kappa\in[\nu/z_{1},\nu]$, where we included also the boundary point $\nu/z_{1}$ for simplicity, it holds that $\tilde{a}^{2}(\pi/2,\nu/\kappa,\kappa)\geq\tilde{a}^{2}(\pi/2,z_{1},\kappa)$ and the equality is saturated for $\kappa=\nu/z_{1}$.
As a consequence,
\begin{eqnarray}\label{Frelation}
h\left(\frac{\pi}{2},\frac{\nu}{\kappa},\kappa\right)&\geq& h\left(\frac{\pi}{2},z_{1},\kappa\right)
\geq h\left(\frac{\pi}{2},z_{1},\frac{\nu}{z_{1}}\right),\nonumber\\
\end{eqnarray}
where the second inequality follows from the fact that the function $\kappa/(\kappa+1)^{2}$ is
monotonically decreasing. Hence we see, that on the interval $\kappa\in[\nu/z_{1},\nu]$ the function
$h\left(\pi/2,\nu/\kappa,\kappa\right)$ attains minimum of
\begin{equation}\label{hmin2}
h_{\rm min}^{(2)}\equiv h\left(\frac{\pi}{2},z_{1},\frac{\nu}{z_{1}}\right)=\frac{49-\sqrt{97}}{128}\doteq 0.3059,
\end{equation}
which is strictly larger than the minimum $h_{\rm min}^{(1)}$, Eq.~(\ref{hmin1}). As a result, the minimum (\ref{hmin1})
coincides with the sought minimal value $h_{\rm min}$, Eq.~(\ref{hmin}), which gives after the substitution into the formula
(\ref{Uboundhom}) the upper bound
\begin{equation}\label{U6}
U\left(\tilde{\rho}_{AB}^{(6)}\right)=\ln\left(\frac{6}{5}\right).
\end{equation}

The results of the present subsection show, that it is possible to calculate the upper bound
(\ref{Ubound}) even for a generic GLEMS which does not possess any further symmetry.
Although we have derived the bound for a single particular state, in the course of the derivation we just used
inequalities $a>b, \tilde{M}<0, z_{1}<\nu$ and $x_{5}x_{7}\nu^{2}+x_{1}x_{3}>0$. Therefore,
for all states $\rho_{AB}^{(6)}$ which are entangled GLEMS satisfying condition (\ref{newGbound})
and the latter inequalities, the quantity $h_{\rm min}$, Eq.~(\ref{hmin}), is equal to $h_{\rm min}=-4M\tilde{M}/(a^2-b^2)^{2}$,
Eq.~(\ref{hmin1}), and the upper bound (\ref{Ubound})
then reads as follows:
\begin{equation}\label{U6generic}
U\left(\rho_{AB}^{(6)}\right)=\ln\left(\frac{a^2-b^2}{\sqrt{D}}\right).
\end{equation}

Note finally, that one can expect that a straightforward modification of previous
procedure would allow us to derive the upper bound (\ref{Ubound}) also for a subclass of states
$\rho_{AB}^{(7)}$, which in addition to condition $a<b$, satisfy inequality $M<0$ as
well as respective analogies of other inequalities needed for derivation of formula (\ref{U6generic}).
While this programme is deferred for further research, in the following section we show,
that for all states investigated in the present section the formulas for the upper bound (\ref{Ubound})
in fact coincide with the GIE.

\subsection{Saturation of the upper bound for GLEMS}\label{subsec_L_GLEMS}

Let us start with observation, that from Eqs.~(\ref{gammaABEgammaE}), (\ref{GLEMSgammaABEgammaE0}) and (\ref{invSgeneric}) it follows that
$\gamma_{E}=\nu\openone$, whereas for blocks $\gamma_{AE}$ and $\gamma_{BE}$ of matrix $\gamma_{ABE}$,
Eq.~(\ref{gammaABgammaAEgammaBE}), one gets
\begin{eqnarray}\label{gammaAEgammaBE}
\gamma_{AE}&=&\sqrt{\nu^{2}-1}\left(\begin{array}{cc}
x_{3} & 0  \\
0 & -x_{1} \\
\end{array}\right),\nonumber\\
\gamma_{BE}&=&\sqrt{\nu^{2}-1}\left(\begin{array}{cc}
x_{4} & 0  \\
0 & -x_{2} \\
\end{array}\right).
\end{eqnarray}
Substituting the latter matrices into Eqs.~(\ref{XAB}) and
(\ref{Xj}), setting $\Gamma_{A}=\Gamma_{A}^{t}=\mbox{diag}(e^{-2t},e^{2t})$ and
$\Gamma_{B}=\Gamma_{B}^{t}=\mbox{diag}(e^{-2t},e^{2t})$, and performing the limit
$t\rightarrow+\infty$, one finds
after some algebra that the matrices $X_{AB}, X_{A}$ and $X_{B}$
attain the same form
\begin{eqnarray}\label{XGLEMS}
X_{k}=\nu\openone-\alpha_{k}|0\rangle\langle0|,
\end{eqnarray}
$k=A,B,AB$, where
\begin{eqnarray}\label{alphaAB}
\alpha_{A}&=&\left(\frac{\nu^{2}-1}{a}\right)x_{3}^{2},\quad
\alpha_{B}=\left(\frac{\nu^{2}-1}{b}\right)x_{4}^{2},\nonumber\\
\alpha_{AB}&=&\left(\frac{\nu^{2}-1}{ab-k_{x}^{2}}\right)(ax_{4}^{2}+bx_{3}^{2}-2k_{x}x_{3}x_{4}),
\end{eqnarray}
and $|0\rangle=(1,0)^{T}$. By substituting from Eq.~(\ref{XGLEMS})
for matrices $X_{A}, X_{B}$ and $X_{AB}$ into the RHS of
Eq.~(\ref{K1}), and using the formula \cite{Henderson_81}
\begin{equation}\label{matdetlem}
\mbox{det}(\mathcal{X}+|c\rangle\langle r|)=\left(1+\langle
r|\mathcal{X}^{-1}|c\rangle\right)\mbox{det}\mathcal{X},
\end{equation}
which is valid for any invertible matrix $\mathcal{X}$, we arrive
at the following simple expression for quantity (\ref{K1}),
\begin{eqnarray}\label{K2}
\mathscr{K}_{h}&=&\frac{(1-\alpha_{A}Q)(1-\alpha_{B}Q)}{(1-\alpha_{AB}Q)},
\end{eqnarray}
where $Q\equiv\langle0|(\Gamma_{E}+\nu\openone)^{-1}|0\rangle$. By
calculating the inverse matrix $(\Gamma_{E}+\nu\openone)^{-1}$, we
further get
\begin{equation}\label{Q}
Q=\frac{(\Gamma_{E})_{22}+\nu}{\mbox{det}(\Gamma_{E}+\nu\openone)}.
\end{equation}
Let us now express CM $\Gamma_{E}$ as in Eq.~(\ref{GammaE}), which can be further rewritten in analogy with Eq.~(\ref{gammaAE})
as
\begin{equation}\label{GammaE2}
\Gamma_{E}= \left(
\begin{array}{cc}
V_{+}+V_{-}\cos(2\varphi)  & V_{-}\sin(2\varphi) \\
V_{-}\sin(2\varphi) & V_{+}-V_{-}\cos(2\varphi) \\
\end{array}
\right)
\end{equation}
with
\begin{eqnarray}\label{Vpm}
V_{+}&\equiv&\frac{V_{x}+V_{p}}{2}=\tau\cosh(2t),\nonumber\\
V_{-}&\equiv&\frac{V_{x}-V_{p}}{2}=\tau\sinh(2t).
\end{eqnarray}
Inserting from here for $(\Gamma_{E})_{22}$ into the RHS of Eq.~(\ref{Q}) and taking into account that
\begin{equation}\label{detGammaE}
\mbox{det}(\Gamma_{E}+\nu\openone)=\tau^2+2\tau\cosh(2t)\nu+\nu^{2},
\end{equation}
 one finds the variable $Q$ appearing on the RHS of Eq.~(\ref{K2}) is equal to 
\begin{equation}\label{Qfinal}
Q=\frac{\tau[\cosh(2t)-\sinh(2t)\cos(2\varphi)]+\nu}{\tau^2+2\tau\cosh(2t)\nu+\nu^{2}}.
\end{equation}

For evaluation of the quantity (\ref{L}) it remains to perform minimization on the RHS of Eq.~(\ref{Kmin}).
This can be done by minimization of the quantity (\ref{K2}), where $Q$ is given in
Eq.~(\ref{Qfinal}) over $\varphi\in[0,\pi)$, $\tau\geq1$ and $t\geq0$. In fact, the minimization
can be greatly simplified. Namely, if we look on the RHS of Eq.~(\ref{K2}) we see, that it is a function of just a single variable $Q$.
Thus, if we find the interval of values in which the variable $Q$ may vary, it is sufficient to
minimize the quantity (\ref{K2}) with respect to the single variable $Q$ on the found interval.
Provided that the minimum lies at some point $Q_{\rm min}$ which can be attained for some admissible
values $\varphi_{\rm min}\in[0,\pi)$, $\tau_{\rm min}\in[1,+\infty)$ and $t_{\rm min}\in[0,+\infty)$,
we get the sought optimized quantity (\ref{Kmin}).

The latter interval is easy to find with the help of the following inequalities:
\begin{equation}\label{Qbounds1}
0<\frac{1}{V_x+\nu}\leq Q \leq \frac{1}{V_p+\nu}<\frac{1}{\nu}.
\end{equation}
Here, the inner inequalities follow from the fact that $Q$ lies between the least eigenvalue $1/(V_x+\nu)$ and the largest eigenvalue
$1/(V_p+\nu)$ of matrix $(\Gamma_{E}+\nu\openone)^{-1}$, which are easy to find using the expression of
CM $\Gamma_{E}$ given in  Eq.~(\ref{GammaE}). The outer inequalities represent the lower and the upper bound on the least
and largest eigenvalue, which is reached in the limit $V_{x}\rightarrow+\infty$ and for $V_{p}=0$, respectively.
Instead of carrying out minimization in Eq.~(\ref{Kmin}), we thus calculate the quantity
\begin{equation}\label{calKmin}
\mathcal{K}_{\rm
min}\equiv\mathop{\mbox{inf}}_{Q\in(0,\frac{1}{\nu})}\mathscr{K}_{h},
\end{equation}
which requires to compare the values of $\mathscr{K}_{h}$ at stationary points lying in the interval
$(0,1/\nu)$ as well as at the boundary points $0$ and $1/\nu$ of the interval and find the least value.

To proceed further with evaluation of the quantity (\ref{calKmin}), we need to know the expression of
parameters $\alpha_{A}, \alpha_{B}$ and $\alpha_{AB}$, Eq.~(\ref{alphaAB}), in terms of parameters
$a,b,k_{x}$ and $k_{p}$. In analogy with previous section, we again analyze each of the
considered types of GLEMS separately.

\subsubsection{GLEMS with $a>b$ and $bk_{x}=ak_{p}$}

Let us now move to calculation of the quantity $L$, Eq.~(\ref{L}), for states $\rho_{AB}^{(4)}$.
By taking from Eq.~(\ref{xclass1}) explicit expressions for parameters $x_{3}$ and $x_{4}$ and substituting
them into Eq.~(\ref{alphaAB}), we arrive after some algebra at
\begin{eqnarray}\label{alphaclass4}
\alpha_{A}=\alpha_{AB}=\frac{\nu^{2}-1}{\nu},\quad \alpha_{B}=\left(\frac{\nu^{2}-1}{\nu}\right)\frac{k_{x}^{2}}{ab},
\end{eqnarray}
where $\nu$ is the symplectic eigenvalue defined in Eq.~(\ref{nu4}).
Owing to equality $\alpha_{A}=\alpha_{AB}$ the quantity to be minimized, Eq.~(\ref{K2}), reduces
to the following simple form:
\begin{eqnarray}\label{Kh4}
\mathscr{K}_{h}=1-\alpha_{B}Q.
\end{eqnarray}
Since $\alpha_{B}>0$ in the present case, previous function is monotonically decreasing function of
$Q$ attaining minimum of
\begin{eqnarray}\label{Kh4min}
\mathcal{K}_{\rm min}=1-\left(\frac{\nu^{2}-1}{\nu^{2}}\right)\frac{k_{x}^{2}}{ab}
\end{eqnarray}
at the boundary point $1/\nu$. The point is reached, e.g., for the measurement with CM (\ref{GammaE}), where
$\varphi=\pi/2,\tau=1$ and in the limit for $t\rightarrow+\infty$, which is homodyne detection of quadrature $x_{E}$
on purifying mode $E$. As a consequence, the quantity (\ref{Kh4min}) coincides with the quantity $\mathscr{K}_{\rm min}$, Eq.~(\ref{Kmin}),
and it gives after substitution into Eq.~(\ref{L}) and some algebra
\begin{equation}\label{LGLEMS4}
L\left(\rho_{AB}^{(4)}\right)=\frac{1}{2}\ln\left(\frac{a^{2}}{a^{2}-k_{x}k_{p}}\right)=\ln\left(\frac{a}{\nu}\right).
\end{equation}
Comparison of the latter quantity with the upper bound (\ref{U4}) reveals, that they are equal. Thus we have shown, that for
the class of states considered here the upper bound on GIE is reached by conditional mutual information (\ref{IABE}) for
distribution of outcomes of homodyne detections of quadratures $x_A$ and $x_{B}$ of modes $A$ and $B$, which is minimized over all
measurements on mode $E$. This implies that for GLEMS with $a>b$ and $bk_{x}=ak_{p}$, which satisfy condition (\ref{newGbound}), GIE is given by
\begin{equation}\label{GIEGLEMS4}
E_{\downarrow}^{G}\left(\rho_{AB}^{(4)}\right)=\ln\left(\frac{a}{\nu}\right).
\end{equation}
In particular, for the state with $a=2\sqrt{2}, b=k_{x}=\sqrt{2}$ and $k_{p}=1/\sqrt{2}$,
formula (\ref{GIEGLEMS4}) yields $E_{\downarrow}^{G}(\rho_{AB}^{(4)})=\ln(2\sqrt{2/7})\doteq0.067$.

\subsubsection{GLEMS with $a<b$ and $ak_{x}=bk_{p}$}

Let us now investigate states $\rho_{AB}^{(5)}$, i.e., GLEMS satisfying conditions $a<b$ and $ak_{x}=bk_{p}$.
From Eq.~(\ref{xclass2}) one finds easily that the parameters (\ref{alphaAB}) read as
\begin{eqnarray}\label{alphaclass5}
\alpha_{A}=\left(\frac{\tilde{\nu}^{2}-1}{\tilde{\nu}}\right)\frac{k_{x}^{2}}{ab},\quad \alpha_{B}=\alpha_{AB}=\frac{\tilde{\nu}^{2}-1}{\tilde{\nu}},
\end{eqnarray}
and hence the quantity (\ref{K2}) boils down to
\begin{eqnarray}\label{Kh5}
\mathscr{K}_{h}=1-\alpha_{A}Q.
\end{eqnarray}
By minimizing the RHS with respect to $Q$ on the interval $(0,1/\tilde{\nu})$ and taking into account inequality $\alpha_{A}>0$
one finds immediately the optimized quantity (\ref{Kmin}) to be
\begin{eqnarray}\label{Kh5min}
\mathscr{K}_{\rm min}=1-\left(\frac{\tilde{\nu}^{2}-1}{\tilde{\nu}^{2}}\right)\frac{k_{x}^{2}}{ab},
\end{eqnarray}
and it is again reached if Eve carries out homodyne detection of quadrature $x_{E}$
on her mode $E$. If we now insert the latter quantity into the RHS of Eq.~(\ref{L}),
we get
\begin{equation}\label{GIEGLEMS5}
L\left(\rho_{AB}^{(5)}\right)=\frac{1}{2}\ln\left(\frac{b^{2}}{b^{2}-k_{x}k_{p}}\right)=\ln\left(\frac{b}{\tilde{\nu}}\right).
\end{equation}
Hence we see again, that the conditional mutual information (\ref{IABE}) for fixed homodyne detections of
quadratures $x_{A}$ and $x_{B}$, which is minimized over all CMs $\Gamma_{E}$, Eq.~(\ref{L}),
saturates the upper bound (\ref{U5}) and thus the GIE for states $\rho_{AB}^{(5)}$ satisfying
condition (\ref{newGbound}) is given by
\begin{equation}\label{GIEGLEMS5}
E_{\downarrow}^{G}\left(\rho_{AB}^{(5)}\right)=\ln\left(\frac{b}{\tilde{\nu}}\right).
\end{equation}
Since the example of a state $\rho_{AB}^{(5)}$ with $b=2\sqrt{2}, a=k_{x}=\sqrt{2}$ and $k_{p}=1/\sqrt{2}$ differs
from previous example just by an exchange of the values of $a$ and $b$ and the same holds also for formulas
(\ref{GIEGLEMS4}) and (\ref{GIEGLEMS5}), we get again $E_{\downarrow}^{G}(\rho_{AB}^{(5)})\doteq0.067$.

\subsubsection{Generic GLEMS}

Like in the case of upper bound (\ref{Ubound}) derived in Subsec.~\ref{sec_U_GLEMS}, also the method of
derivation of the quantity (\ref{L}) presented in previous two subsubsections can be extended to more
generic GLEMS. Although we again illustrate this on one concrete state investigated in
Subsubsec.~\ref{subsubsec_U_generic_GLEMS}, we first carry out the derivation in full generality
and the concrete values of the parameters $a,b, k_{x}$ and $k_{p}$ are substituted into
the final formulas only at the end of our calculations. This implies, that most of
the results presented here do not hold only for the considered state but they can be
straightforwardly used to derive the quantity (\ref{L}) also for other generic GLEMS.

In the general case GLEMS with $a>b$ ($a<b$) satisfy $bk_{x}\ne ak_{p}$ ($ak_{x}\ne bk_{p}$),
and the function (\ref{K2}) has two generally different stationary points. They can be obtained
as solutions of the extremal equation $d\mathscr{K}_{h}/dQ=0$, which is equivalent
with the following quadratic equation:
\begin{equation}\label{extremalequation}
\alpha_{A}\alpha_{B}\alpha_{AB}Q^{2}-2\alpha_{A}\alpha_{B}Q+\alpha_{A}+\alpha_{B}-\alpha_{AB}=0,
\end{equation}
and which possesses the following two solutions:
\begin{equation}\label{Q12}
Q_{1,2}=\frac{1}{\alpha_{AB}}\left[1\pm\sqrt{\left(\frac{\alpha_{AB}}{\alpha_{A}}-1\right)\left(\frac{\alpha_{AB}}{\alpha_{B}}-1\right)}\right].
\end{equation}
Using formulas (\ref{alphaAB}) we can now write the ratios in the round brackets
on the RHS of Eq.~(\ref{Q12}) as
\begin{eqnarray}\label{alphafractions}
\frac{\alpha_{AB}}{\alpha_{A}}=1+x^{2},\quad \frac{\alpha_{AB}}{\alpha_{B}}=1+y^{2},
\end{eqnarray}
where we introduced
\begin{eqnarray}\label{xyforQ}
x&\equiv&\frac{a}{\sqrt{ab-k_{x}^2}}\left(\frac{x_{4}}{x_{3}}-\frac{k_{x}}{a}\right),\nonumber\\
y&\equiv&\frac{b}{\sqrt{ab-k_{x}^2}}\left(\frac{x_{3}}{x_{4}}-\frac{k_{x}}{b}\right).
\end{eqnarray}
For sates with $a>b$ and $bk_{x}\ne ak_{p}$ ($a<b$ and $ak_{x}\ne bk_{p}$) we can further substitute here
from relation $x_{4}/x_{3}=-M/L_{1}$, Eq.~(\ref{2b}) of Appendix~\ref{Sdiagonalization} ($x_{4}/x_{3}=\tilde{M}/L_{2}$,
Eq.~(\ref{tildex}) of Appendix~\ref{Sdiagonalization}), where $L_{1}$ and $L_{2}$ are quantities defined
in Eq.~(\ref{ML12}) of Appendix~\ref{Sdiagonalization}. This allows us to express parameters $x$ and $y$, and hence
also the stationary points $Q_{1}$ and $Q_{2}$, in terms of parameters $a, b, k_{x}$ and $k_{p}$.

By introducing parameters $x,y$ we simplified stationary points (\ref{Q12})
to $Q_{1,2}=(1\pm|xy|)/\alpha_{AB}$. With the help of the latter formulas together with
Eq.~(\ref{alphafractions}) we finally express the value of quantity
(\ref{K2}) in the stationary points as
\begin{equation}\label{KQ12}
\mathscr{K}_{h}(Q_{1,2})=\frac{(|x|\mp|y|)^{2}}{(1+x^2)(1+y^2)}.
\end{equation}

To get the optimized quantity (\ref{calKmin}) we now have to identify which of the stationary points $Q_{1,2}$ lies in the interval $(0,1/\nu)$.
By comparing the values (\ref{KQ12}) for the stationary points lying in the interval with the
values of the quantity (\ref{K2}) in the boundary points, $\mathscr{K}_{h}(0)=1$ and
\begin{eqnarray}\label{Khboundary}
\mathscr{K}_{h}\left(\frac{1}{\nu}\right)=\frac{(\nu-\alpha_{A})(\nu-\alpha_{B})}{\nu(\nu-\alpha_{AB})},
\end{eqnarray}
and selecting the least value, we get the quantity (\ref{calKmin}). If one can further find parameters $\varphi, \tau$ and
$t$ of CM $\Gamma_{E}$ giving the value of $Q$ at which the least value is reached, the quantity (\ref{calKmin})
coincides with the quantity (\ref{Kmin}) we are looking for.

By applying previous algorithm to the state $\tilde{\rho}_{AB}^{(6)}$ of Subsubsec.~\ref{subsubsec_U_generic_GLEMS} with $a=2\sqrt{2}, b=\sqrt{2}$
and $k_{x,p}=(\sqrt{97}\pm1)/8$ we get
\begin{eqnarray}\label{Qexample}
Q_{1}\doteq0.578>\frac{1}{\nu}\doteq0.408>Q_{2}\doteq0.402
\end{eqnarray}
and
\begin{eqnarray}\label{Kexample}
\mathscr{K}_{h}(Q_{2})=\frac{9}{800}(79-\sqrt{97})\doteq 0.7780,\nonumber\\
\mathscr{K}_{h}\left(\frac{1}{\nu}\right)=\frac{3169-79\sqrt{97}}{3072}\doteq 0.7783,
\end{eqnarray}
which implies that $\mathcal{K}_{\rm min}=\mathscr{K}_{h}(Q_{2})$. Further, the value of $Q_{2}$ can be reached by a CM $\Gamma_{E}$,
Eq.~(\ref{GammaE}), with $\varphi=\pi/2$, $\tau=1$ and $t\doteq 1.613$, which corresponds to a projection
onto a pure state squeezed in quadrature $x_{E}$ with finite squeezing. Consequently, it holds that $\mathscr{K}_{\rm min}=\mathscr{K}_{h}(Q_{2})$,
the quantity (\ref{L}) is equal to $L(\tilde{\rho}_{AB}^{(6)})=\ln(6/5)$ and it coincides with the upper
bound (\ref{U6}). Hence, GIE for the state $\tilde{\rho}_{AB}^{(6)}$ is given by
\begin{equation}\label{GIEGLEMS6}
E_{\downarrow}^{G}\left(\tilde{\rho}_{AB}^{(6)}\right)=\ln\left(\frac{6}{5}\right).
\end{equation}

Needles to say finally, that the quantity $L$, Eq.~(\ref{L}), represents at fixed homodyne detections
of quadratures $x_{A}$ and $x_{B}$ on modes $A$ and $B$ the least conditional mutual information over all Gaussian
measurements on mode $E$, and thus it gives a lower bound on GIE. In the course of derivation of GIE for
symmetric GLEMS and asymmetric squeezed thermal GLEMS performed in Ref.~\cite{Mista_16}, equality of the upper
bound (\ref{Ubound}) given by the RHS of Eqs.~(\ref{GIEsymGLEMS}) and (\ref{GIEasymGLEMS}) to the latter
minimal conditional mutual information was utilized, without proving it explicitly. In Appendix~\ref{Lcalculation}
we prove the equality by showing equality of the quantity $L$ to the RHS of formulas (\ref{GIEsymGLEMS}) and (\ref{GIEasymGLEMS}),
thereby complementing derivation of GIE for symmetric GLEMS and asymmetric squeezed thermal GLEMS.

\section{RELATION TO OTHER ENTANGLEMENT MEASURES}\label{sec_Relation_to_other_measures}

An important question which has to be addressed for any entanglement quantifier is its relation to other
know entanglement quantifiers. Here, we first study whether there is a connection between GIE and
the most popular easily computable logarithmic negativity \cite{Vidal_02,Eisert_PhD,Plenio_05}. Next,
we move to analysis of relation of GIE and the GR2EoF \cite{Adesso_12,Lami_17}.

\subsection{Relation to logarithmic negativity}

The obtained formulas for GIE of symmetric states can be compactly
written as \cite{Mista_16}
\begin{equation}\label{GIEsym}
E_{\downarrow}^{G}(\rho_{AB})=\left\{\begin{array}{lll} \ln\left[\frac{\tilde{\nu}_{-}+(\tilde{\nu}_{-})^{-1}}{2}\right], & \textrm{if} & \tilde{\nu}_{-}<1;\\
0, & \textrm{if} & \tilde{\nu}_{-}\geq1,
\end{array}\right.
\end{equation}
where $\tilde{\nu}_{-}=\sqrt{(a-k_{x})(a-k_{p})}$ is the lower
symplectic eigenvalue of the partial transpose of the investigated state
$\rho_{AB}$. Hence we see, that for considered symmetric states
the GIE is a monotonically decreasing function of the symplectic
eigenvalue $\tilde{\nu}_{-}$ and thus in this respect it stays in
line with other frequently used Gaussian entanglement measure
called logarithmic negativity \cite{Vidal_02,Eisert_PhD,Plenio_05}
which is defined as $E_{\cal{N}}(\rho_{AB})=\mbox{max}[0,-\ln\tilde{\nu}_{-}]$.
However, for asymmetric states this rule is violated as it is
witnessed by the formula (\ref{GIEasymGLEMS}), which cannot be
rewritten as a function solely of the symplectic eigenvalue
$\tilde{\nu}_{-}=[a+b-\sqrt{(a+b)^2-4\nu}]/2$, where $\nu=1+|a-b|$.

Further comparison of GIE with logarithmic negativity unveils that
$E_{\cal{N}}(\rho_{AB})\geq E_{\downarrow}^{G}(\rho_{AB})$ in the case
of symmetric states and thus one might be tempted to think,
that there is a fixed hierarchy between the two quantifiers which holds for
all bipartite Gaussian states. While this conjecture might be true
for two-mode Gaussian states, it is false in general because
$E_{\cal{N}}(\rho_{AB}^{\rm PPT})=0$ for Gaussian entangled states
with positive partial transpose ($\equiv\rho_{AB}^{\rm PPT}$) \cite{Werner_01},
whereas $E_{\downarrow}^{G}(\rho_{AB}^{\rm PPT})>0$ due to the
faithfulness property.
\subsection{Relation to Gaussian R\'{e}nyi-2 entanglement of formation}

A systematic investigation of GIE carried out in
Refs.~\cite{Mista_15,Mista_16} revealed a remarkable relation of
the quantity to GR2EoF \cite{Adesso_12,Lami_17}. Concretely, it was shown,
that formulas for GIE for symmetric GLEMS, Eq.~(\ref{GIEsymGLEMS}), symmetric squeezed thermal states,
Eq.~(\ref{GIEsymTMST}), as well as asymmetric squeezed thermal GLEMS, Eq.~(\ref{GIEasymGLEMS}),
{\it coincide} with the formulas for GR2EoF. Based on this observation, in Ref.~\cite{Mista_16} a conjecture
has been expressed that GIE and GR2EoF are equal on all bipartite Gaussian states.

In this section we further strengthen the conjecture by showing, that also expressions
(\ref{GIEGLEMS4}), (\ref{GIEGLEMS5}) and (\ref{GIEGLEMS6}) for GIE of GLEMS $\rho_{AB}^{(4)}$,
$\rho_{AB}^{(5)}$, as well as an example state $\tilde{\rho}_{AB}^{(6)}$,
are again equal to GR2EoF.

Recall, that GR2EoF ($\equiv E_{F,2}^{G}$) for a two-mode reduction
of a pure state of three modes $A_{1}, A_{2}$ and $A_{3}$ can be
calculated from the following standard-form CM of the state \cite{Adesso_06}:
\begin{eqnarray}\label{threemodepure}
\gamma_{A_{1}A_{2}A_{3}}=\left(\begin{array}{cccccc}
a_{1} & 0 & c_{3}^{+} & 0 & c_{2}^{+} & 0\\
0 & a_{1} & 0 & c_{3}^{-} & 0 & c_{2}^{-}\\
c_{3}^{+} & 0 & a_{2} & 0 & c_{1}^{+} & 0\\
0 & c_{3}^{-} & 0 & a_{2} & 0 & c_{1}^{-}\\
c_{2}^{+} & 0 & c_{1}^{+} & 0 & a_{3} & 0\\
0 & c_{2}^{-} & 0 & c_{1}^{-} & 0 & a_{3}\\
\end{array}\right),
\end{eqnarray}
with
\begin{eqnarray}\label{ci}
c_{i}^{\pm}&=&\frac{\sqrt{a_{--}a_{+-}}\pm\sqrt{a_{-+}a_{++}}}{4\sqrt{a_{j}a_{k}}},
\end{eqnarray}
where
\begin{eqnarray}\label{apm}
a_{\mp-}&=&(a_{i}\mp1)^2-(a_{j}-a_{k})^{2},\nonumber\\
a_{\mp+}&=&(a_{i}\mp1)^2-(a_{j}+a_{k})^{2},
\end{eqnarray}
and $|a_{j}-a_{k}|+1\leq a_{i}\leq a_{j}+a_{k}-1$, where
$\{i,j,k\}$ run over all possible permutations of $\{1,2,3\}$.
The GR2EoF of a reduced state $\rho_{A_{i}A_{j}}$ of
modes $A_{i}$ and $A_{j}$ with CM $\gamma_{A_{i}A_{j}}$ is given by
\cite{Adesso_12}
\begin{equation}\label{GR2EoF}
E_{F,2}^{G}\left(\rho_{A_{i}A_{j}}\right)=\frac{1}{2}\ln g_{k},
\end{equation}
where \cite{Adesso_05}
\begin{equation}\label{gk}
g_{k}=\left\{\begin{array}{lll} 1, & \textrm{if} & a_{k}\geq\sqrt{a_{i}^{2}+a_{j}^{2}-1};\\
\frac{\zeta}{8a_{k}^{2}}, & \textrm{if} & \alpha_{k}<a_{k}<\sqrt{a_{i}^{2}+a_{j}^{2}-1};\\
\left(\frac{a_{i}^2-a_{j}^2}{a_{k}^2-1}\right)^{2}, & \textrm{if}
& a_{k}\leq\alpha_{k}.
\end{array}\right.
\end{equation}
Here,
\begin{eqnarray}\label{alphakdeltazeta}
\alpha_{k}&=&\left\{1+\frac{(a_{i}^{2}-a_{j}^{2})^{2}}{2(a_{i}^{2}+a_{j}^{2})}+\frac{|a_{i}^{2}-a_{j}^{2}|}{2(a_{i}^{2}+a_{j}^{2})}\right.\nonumber\\
&&\left.\times\left[(a_{i}^{2}-a_{j}^{2})^{2}+8(a_{i}^{2}+a_{j}^{2})\right]^{\frac{1}{2}}\right\}^{\frac{1}{2}},\nonumber\\
\delta&=&[(a_{1}-a_{2}-a_{3})^{2}-1][(a_{1}+a_{2}-a_{3})^{2}-1]\nonumber\\
&&\times[(a_{1}-a_{2}+a_{3})^{2}-1][(a_{1}+a_{2}+a_{3})^{2}-1],\nonumber\\
\zeta&=&2a_{1}^2+2a_{2}^2+2a_{3}^2+2a_{1}^{2}a_{2}^{2}+2a_{1}^{2}a_{3}^{2}+2a_{2}^{2}a_{3}^{2}\nonumber\\
&&-a_{1}^{4}-a_{2}^{4}-a_{3}^{4}-\sqrt{\delta}-1.
\end{eqnarray}

At the outset, we calculate GR2EoF for states $\rho_{AB}^{(4)}$. In this case, from Eqs.~(\ref{gammast2}), (\ref{gammapi})
and the fact that $\gamma_{E}=\nu\openone$, we get immediately the parameters of the pure-state CM
(\ref{threemodepure}), which are needed for evaluation of GR2EoF, in the form $a_{1}=a, a_{2}=b$ and $a_{3}=\nu$. According to
Eq.~(\ref{GR2EoF}) the GR2EoF is equal to $E_{F,2}^{G}\left(\rho_{AB}^{(4)}\right)=\ln(g_{3})/2$, where $g_{3}$ is obtained from
Eq.~(\ref{gk}) for $k=3$, and where we identified modes as $A\equiv A_{1}, B\equiv A_{2}$ and $E\equiv A_{3}$. Since $\sqrt{a_{1}^{2}+a_{2}^{2}-1}=\sqrt{a^2+k_{x}k_{p}}>\sqrt{a^2-k_{x}k_{p}}=\nu=a_{3}$,
where the first equality follows from condition $\nu_{2}=\sqrt{b^2-k_{x}k_{p}}=1$ and the inequality is a consequence of the
assumption $k_{x}k_{p}>0$, the first branch of Eq.~(\ref{gk}) never applies. The latter two conditions further
imply, that $b>1$ and thus $(a^2+b^2)(b^2-1)>0$ which is equivalent with inequality
\begin{equation}\label{ineq1}
\frac{a^2-b^2+\sqrt{(a^2-b^2)^2+8(a^2+b^2)}}{2(a^2+b^2)}<1
\end{equation}
being further equivalent with inequality $\alpha_{3}<a_{3}$, and thus only second branch in Eq.~(\ref{gk}) applies. Next, after
some algebra one finds that $\sqrt{\delta}=4(a^2-b^2)(b^2-1)$ and consequently $\zeta=8a^2$, where the condition $a>b$ has been used.
As a result, the GR2EoF for states $\rho_{AB}^{(4)}$ is equal to
\begin{equation}\label{GR2EoF4}
E_{F,2}^{G}\left(\rho_{AB}^{(4)}\right)=\ln\left(\frac{a}{\nu}\right).
\end{equation}
By comparing the latter formula with GIE for the same class of states, Eq.~(\ref{GIEGLEMS4}), we see that
like in many other cases the two quantities coincide.

Derivation of GR2EoF for the class of states $\rho_{AB}^{(5)}$ can be performed exactly in the same way as in the previous
case. As for parameters of CM (\ref{threemodepure}) one has $a_{1}=a, a_{2}=b$ and $a_{3}=\tilde{\nu}$ and condition
$\nu_{2}=\sqrt{a^2-k_{x}k_{p}}=1$ is obeyed, it is obvious that $\sqrt{a_{1}^{2}+a_{2}^{2}-1}=\sqrt{b^2+k_{x}k_{p}}>\sqrt{b^2-k_{x}k_{p}}=\tilde{\nu}=a_{3}$
and the first branch of Eq.~(\ref{gk}) never occurs. Additionally, since inequality obtained by exchanging $a$ and $b$ in inequality
(\ref{ineq1}) is fulfilled for states $\rho_{AB}^{(5)}$, we have $\alpha_{3}<a_{3}$ for the states and thus one always has to take the
second branch of Eq.~(\ref{gk}). Finally, because $\sqrt{\delta}=4(b^2-a^2)(a^2-1)$ and $\zeta=8b^2$, the GR2EoF is equal to
\begin{equation}\label{GR2EoF5}
E_{F,2}^{G}\left(\rho_{AB}^{(5)}\right)=\ln\left(\frac{b}{\tilde{\nu}}\right)
\end{equation}
according to Eq.~(\ref{GR2EoF}). A quick look at formula (\ref{GIEGLEMS5}) uncovers again
that equality of GIE and GR2EoF holds also for the class of states $\rho_{AB}^{(5)}$.

We conclude this section by calculating the GR2EoF for the state $\tilde{\rho}_{AB}^{(6)}$ from Subsubsection~\ref{subsubsec_U_generic_GLEMS}. In this case, one has $a=2\sqrt{2}$, $b=\sqrt{2}$ and $k_{x,p}=(\sqrt{97}\pm1)/8$, whence the relevant parameters of CM~(\ref{threemodepure}) are given by $a_1=a$, $a_2=b$ and $a_3=\nu=\sqrt{6}\doteq 2.45$. Further, one finds out that $a_3<\alpha_3=\sqrt{(14+3\sqrt{29})/5}\doteq 2.456$ and thus the third branch of Eq.~(\ref{gk}) has to be taken. Finally, according to Eq.~(\ref{GR2EoF}) the GR2EoF is equal to
\begin{equation}
E_{F,2}^G\left(\tilde{\rho}_{AB}^{(6)}\right) = \ln\left(\frac{6}{5}\right).
\end{equation}
Comparing the result with Eq.~(\ref{GIEGLEMS6}) one can see that GIE and GR2EoF are equal also
for this state.

The results presented in this subsection reveal, that also in the case of states $\rho_{AB}^{(4)}$, $\rho_{AB}^{(5)}$
as well as $\tilde{\rho}_{AB}^{(6)}$ GIE coincides with GR2EoF. This even strengthens already a strongly supported
conjecture about equivalence of the two quantities.

\section{CONCLUSIONS}\label{sec_Conclusions}

Entanglement quantification based on tripartite extensions of quantified
states as embodied by squashed entanglement \cite{Christandl_04} proves to be
currently most successful. This is not only because this approach allows to
fulfil all axioms imposed on a good entanglement measure \cite{Brandao_11}, but
also due to the fact that the cryptographic nature of the underlying scenario
may give the quantifier a cryptographic meaning \cite{Takeoka_14}.

In this paper, we developed a theory of another representative of such quantifiers
called GIE, which was initiated in Refs.~\cite{Mista_15,Mista_16}. The GIE is in fact a
Gaussian relative of squashed entanglement because it is defined as optimized intrinsic information
being the mother of squashed entanglement.

First, we have shown, that the analytic formulas for GIE derived in Ref.~\cite{Mista_16} hold for larger
classes of states than previously thought. Second, we have derived analytical expressions for
GIE for two new classes of two-mode Gaussian states with partial minimum uncertainty, which have
certain symmetry. Moreover, by deriving GIE for one particular state we have demonstrated,
that it can be calculated also for generic partial minimum-uncertainty states which possess no
further symmetry. Finally, we have proved, that like for all states studied in Ref.~\cite{Mista_16},
also for the new states investigated here the GIE is always equal to the GR2EoF. In view of our results
we think, that equality of GIE and GR2EoF conjectured in Ref.~\cite{Mista_16} is very likely. The proof of this
conjecture, which would equip Gaussian entanglement theory with a unique entanglement measure, is left
for further research.

\acknowledgments{L. M. would like to thank G. Adesso and L. Lami for
fruitful discussions.}

\appendix
\section{Symplectic diagonalization}\label{Sdiagonalization}

In this section we derive for any CM (\ref{gammast2}) an explicit
expression for a symplectic matrix $S$ which brings the CM to the
Williamson normal form (\ref{Williamson}). This can be done either
using a method of Ref.~\cite{Serafini_05} or a method of
Ref.~\cite{Pirandola_09}. In the first method \cite{Serafini_05}
we seek matrix $S$ in the form of a product
$S=\left(\oplus_{i=1}^{2}U^{\ast}\right)V^{T}$, where
\begin{equation}\label{U}
U=\frac{1}{\sqrt{2}}\left(\begin{array}{cc}
i & -i \\
1 & 1\\
\end{array}\right)
\end{equation}
and $V$ contains in its columns the eigenvectors of the matrix
$i\Omega\gamma_{AB}$, which are chosen such that $S$ is real and
it satisfies the symplectic condition (\ref{Scondition}) with
$N=2$, $S\Omega_{2}S^{T}=\Omega_{2}$. The matrix $S$ can be found
using the aforementioned method with an additional constraint that
it does not mix position and momentum quadratures. Hence we get
\begin{eqnarray}\label{S}
S=\left(\begin{array}{cccc}
x_1 & 0 & x_2 & 0\\
0 & x_3 & 0 & x_4 \\
x_5 & 0 & x_6 & 0 \\
0 & x_7 & 0 & x_8 \\
\end{array}\right),
\end{eqnarray}
where the real parameters $x_1,x_{2},\ldots,x_8$ are related to the
elements of eigenvectors $u_{\nu_{1}}$ and $w_{\nu_{2}}$ of the
matrix $i\Omega\gamma_{AB}$ corresponding to the eigenvalues
$\nu_{1}$ and $\nu_{2}$ as
$u_{\nu_{1}}=\left(ix_1,x_3,ix_2,x_4\right)^{T}$ and
$w_{\nu_{2}}=\left(ix_5,x_7,ix_6,x_8\right)^{T}$ and thus satisfy
the set of equations
\begin{eqnarray}\label{constraints1}
M x_3+L_{1}x_4=0,\quad M x_7+L_{2}x_8=0,\nonumber\\
\tilde{M} x_4-L_{2}x_3=0,\quad \tilde{M} x_8-L_{1}x_7=0,\nonumber\\
x_1=\frac{ax_3-k_px_4}{\nu_{1}},\quad x_2=\frac{bx_4-k_px_3}{\nu_{1}},\nonumber\\
x_5=\frac{ax_7-k_px_8}{\nu_{2}},\quad
x_6=\frac{bx_8-k_px_7}{\nu_{2}},
\end{eqnarray}
where we defined
\begin{eqnarray}\label{ML12}
M\equiv ak_x-bk_p,\quad \tilde{M}\equiv bk_x-ak_p,\nonumber\\
L_{1,2}\equiv b^2-k_xk_p-\nu_{1,2}^2=\frac{b^2-a^2\mp\sqrt{D}}{2},
\end{eqnarray}
where quantity $D$ is defined in Eq.~(\ref{DeltaD}).
On the top of that, parameters $x_1,\ldots,x_8$ also have to
satisfy further constraints
\begin{eqnarray}\label{constraints2}
x_1x_3+x_2x_4&=&1,\quad x_1x_7+x_2x_8=0,\nonumber\\
x_5x_7+x_6x_8&=&1,\quad x_3x_5+x_4x_6=0
\end{eqnarray}
imposed by condition $S\Omega_{2}S^{T}=\Omega_{2}$.

Several cases must be distinguished when solving sets of equations
(\ref{constraints1}) and (\ref{constraints2}) depending on the
relations between the parameters $a,b,k_x$ and $k_p$.

1. If $a=b$ and $k_x\geq k_p>0$, Eq.~(\ref{nu12}) gives
\begin{eqnarray}\label{nu12sym}
\nu_{1}&=&\sqrt{\left(a+k_x\right)\left(a-k_p\right)},\nonumber\\
\nu_{2}&=&\sqrt{\left(a-k_x\right)\left(a+k_p\right)},
\end{eqnarray}
and CM (\ref{gammast2}) is brought to the Williamson normal form
(\ref{Williamson}) by symplectic matrix
\begin{eqnarray}\label{S2}
S_{1}=\frac{1}{\sqrt{2}}\left(\begin{array}{cccc}
z_{A}^{-1} & 0 & z_{A}^{-1} & 0\\
0 & z_{A} & 0 & z_{A} \\
-z_{B} & 0 & z_{B} & 0 \\
0 & -z_{B}^{-1} & 0 & z_{B}^{-1} \\
\end{array}\right),
\end{eqnarray}
where $z_{A}=\sqrt[4]{\frac{a+k_x}{a-k_p}}>1$ and
$z_{B}=\sqrt[4]{\frac{a+k_p}{a-k_x}}>1$. A closer look at matrix
(\ref{S2}) reveals that it can be expressed as the following
product
\begin{equation}\label{Ssym}
S=(S_{A}\oplus S_{B})U_{BS},
\end{equation}
which describes a composition of a balanced beam splitter
described by symplectic matrix
\begin{equation}\label{Ubeamsplitter}
U_{BS}=\frac{1}{\sqrt{2}}\left(\begin{array}{cc}
\openone & \openone \\
-\openone & \openone\\
\end{array}\right)
\end{equation}
followed by local squeezing transformations of quadratures $x_{A}$
and $p_{B}$, described by diagonal symplectic matrices
\begin{equation}\label{SAB}
S_{A}=\left(\begin{array}{cc}
z_{A}^{-1} & 0 \\
0 & z_{A}\\
\end{array}\right),\quad S_{B}=\left(\begin{array}{cc}
z_{B} & 0 \\
0 & z_{B}^{-1}\\
\end{array}\right).
\end{equation}

2. If $a>b$, then $M>0$ and $L_{1}<0$.

a) If $bk_x=ak_p$, we get $\tilde{M}=0$ and
\begin{eqnarray}\label{nuAnuB2a}
\nu_{1}=\sqrt{a^2-k_xk_p},\quad \nu_{2}=\sqrt{b^2-k_xk_p}
\end{eqnarray}
by Eq.~(\ref{nu12}). Further, $L_{2}=0$ and by solving Eqs.~(\ref{constraints1}) and
(\ref{constraints2}) we arrive at matrix (\ref{S}) of the form
\begin{eqnarray}\label{S2a}
S_{2a}=\left(\begin{array}{cccc}
\sqrt{\frac{\nu_{1}}{a}} & 0 & 0 & 0\\
0 & \sqrt{\frac{a}{\nu_{1}}} & 0 & \frac{k_x}{\sqrt{a\nu_{1}}} \\
\frac{-k_p}{\sqrt{b\nu_{2}}} & 0 & \sqrt{\frac{b}{\nu_{2}}} & 0 \\
0 & 0 & 0 &  \sqrt{\frac{\nu_{2}}{b}}\\
\end{array}\right).
\end{eqnarray}
Similar to case 1. we can decompose the symplectic matrix in terms of more simple symplectic matrices.
Here, the decomposition attains the following form:
\begin{equation}\label{S2adecomposition}
S_{2a}=(S_{A}\oplus S_{B})S_{QND},
\end{equation}
where the matrices $S_{A}$ and $S_{B}$ are given in Eq.~(\ref{SAB}) with
$z_{A}=\sqrt{\frac{a}{\nu_{1}}}>1$ and $z_{B}=\sqrt{\frac{b}{\nu_{2}}}>1$, and
they describe squeezing in quadratures $x_{A}$ and $p_{B}$ of modes $A$ and $B$,
respectively. The matrix $S_{QND}$ is a symplectic matrix of the quantum non-demolishing
interaction
\begin{eqnarray}\label{SQND}
S_{QND}=\left(\begin{array}{cccc}
1 & 0 & 0 & 0\\
0 & 1 & 0 & q \\
-q & 0 & 1 & 0 \\
0 & 0 & 0 &  1\\
\end{array}\right),
\end{eqnarray}
with interaction constant $q=\frac{k_x}{a}=\frac{k_{p}}{b}$.

Note, that the present set of states is non-empty. For instance, a CM
(\ref{gammast2}) with $a=3,b=2,k_x=2$ and $k_p=4/3$ satisfies both inequalities (\ref{CMcondition}) and thus
represents a physical state with $a>b$ and $ak_p=bk_x$.

b) If $bk_x>ak_p$ we have $L_{2}>0$, whereas if $bk_x<ak_p$ we have
$L_{2}<0$. Thus, if $bk_x\ne ak_p$ then $L_{2}\ne 0$. By solving
Eqs.~(\ref{constraints1}) and (\ref{constraints2}) we get a matrix
$S_{2b}$ of the form (\ref{S}) with
\begin{eqnarray}\label{2b}
x_3&=&-\frac{L_{1}}{M}x_{4},\quad x_7=-\frac{L_{2}}{M}x_{8},\nonumber\\
x_1&=&-\frac{aL_{1}+k_pM}{\nu_{1} M}x_4,\quad x_2=\frac{k_pL_{1}+bM}{\nu_{1} M}x_4,\nonumber\\
x_5&=&-\frac{aL_{2}+k_pM}{\nu_{2} M}x_8,\quad x_6=\frac{k_pL_{2}+bM}{\nu_{2} M}x_8,\nonumber\\
x_4&=&M\sqrt{\frac{\nu_{1}}{aL_{1}^2+2k_pL_{1}M+bM^2}},\nonumber\\
x_8&=&M\sqrt{\frac{\nu_{2}}{aL_{2}^2+2k_pL_{2}M+bM^2}}.
\end{eqnarray}

3. If $a<b$, we can find the symplectic matrix (\ref{S}) by transforming the present case to case 2
with the help of symplectic matrix
\begin{equation}\label{T}
T=\left(\begin{array}{cc}
0 & \openone \\
\openone & 0\\
\end{array}\right),
\end{equation}
which exchanges modes $A$ and $B$. The matrix $T$ transforms CM $\gamma_{AB}$, Eq.~(\ref{gammast2}), to CM
$\tilde{\gamma}_{AB}\equiv T\gamma_{AB}T^{T}$, which is again of the form (\ref{gammast2}), but with parameters
$a$ and $b$ exchanged. Next, we calculate a symplectic matrix ($\equiv \tilde{S}$) which brings CM $\tilde{\gamma}_{AB}$
to the Williamson normal form by solving a set of equations obtained from Eq.~(\ref{constraints1}) by exchanging
parameters $a$ and $b$, and a set of equations (\ref{constraints2}). The symplectic matrix which
brings the original CM (\ref{gammast2}) to the Williamson normal form is then given by $S=\tilde{S}T$.

Let us now apply previous algorithm do derive an explicit form of the symplectic matrix $S$, Eq.~(\ref{S}), which brings
CM (\ref{gammast2}) with $a<b$ to the Williamson normal form. We have already mentioned, that one set of
equations to be solved to get $S$ is obtained from Eqs.~(\ref{constraints1}) by exchanging parameters $a$ and $b$.
This entails the following exchanges $M\leftrightarrow \tilde{M}$ and $L_{1,2}\rightarrow -L_{2,1}$, whereas
symplectic eigenvalues (\ref{nu12}) remain unchanged. In analogy with previous case 2 we see, that
if $a<b$ then $\tilde{M}>0$, $L_{2}>0$, and in
dependence on whether the quantity $M$ vanishes or not, we distinguish the following two cases:

a) If $ak_{x}=bk_{p}$, we get $M=0$ and hence
\begin{eqnarray}\label{nuAnuB3a}
\nu_{1}=\sqrt{b^2-k_xk_p},\quad \nu_{2}=\sqrt{a^2-k_xk_p},
\end{eqnarray}
as well as $L_{1}=0$. The symplectic matrix $\tilde{S}_{3a}$ is obtained from the symplectic matrix
(\ref{S2a}) by exchanging $a$ and $b$. By multiplying the latter matrix by the matrix (\ref{T}) from the right,
we finally get
\begin{eqnarray}\label{S3a}
S_{3a}=\left(\begin{array}{cccc}
0 & 0 & \sqrt{\frac{\nu_{1}}{b}} & 0\\
0 & \frac{k_x}{\sqrt{b\nu_{1}}} & 0 & \sqrt{\frac{b}{\nu_{1}}}\\
\sqrt{\frac{a}{\nu_{2}}} & 0 &  \frac{-k_p}{\sqrt{a\nu_{2}}} & 0\\
0 & \sqrt{\frac{\nu_{2}}{a}} & 0 &  0\\
\end{array}\right).
\end{eqnarray}

b) If $ak_x>bk_p$ we have $L_{1}<0$, while if $ak_x<bk_p$ we have
$L_{1}>0$. Thus, if $ak_x\ne bk_p$ then $L_{1}\ne0$. By solving
Eqs.~(\ref{constraints1}) and (\ref{constraints2}) with $a$ and $b$ exchanged we get a matrix
$\tilde{S}_{3b}$ of the form (\ref{S}) with
\begin{eqnarray}\label{tildex}
x_3&=&\frac{L_{2}}{\tilde{M}}x_{4},\quad x_{7}=\frac{L_{1}}{\tilde{M}}x_{8},\nonumber\\
x_1&=&\frac{b L_{2}-k_p\tilde{M}}{\nu_{1} \tilde{M}}x_4,\quad x_2=\frac{a\tilde{M}-k_p L_{2}}{\nu_{1}\tilde{M}}x_4,\nonumber\\
x_5&=&\frac{b L_{1}-k_p\tilde{M}}{\nu_{2} \tilde{M}}x_8,\quad x_6=\frac{a\tilde{M}-k_p L_{1}}{\nu_{2} \tilde{M}}x_8,\nonumber\\
x_4&=&\tilde{M}\sqrt{\frac{\nu_{1}}{b L_{2}^2-2k_p L_{2}\tilde{M}+a\tilde{M}^2}},\nonumber\\
x_8&=&\tilde{M}\sqrt{\frac{\nu_{2}}{b L_{1}^2-2k_p L_{1}\tilde{M}+a\tilde{M}^2}}.
\end{eqnarray}
Hence, the sought matrix $(\equiv S_{3b})$, which brings the original CM (\ref{gammast2}) with $a<b$ and
$ak_x\ne bk_p$ to the Williamson normal form reads as
\begin{eqnarray}\label{S3b}
S_{3b}=\tilde{S}_{3b}T=\left(\begin{array}{cccc}
x_2 & 0 & x_1 & 0\\
0 & x_4 & 0 & x_3 \\
x_6 & 0 & x_5 & 0 \\
0 & x_8 & 0 & x_7 \\
\end{array}\right),
\end{eqnarray}
where the elements $x_{1}, x_{2},\ldots,x_{8}$ are given in Eq.~(\ref{tildex}).

Needles to say finally, that sets of Eqs.~(\ref{constraints1}) and (\ref{constraints2}), which we used to derive symplectic
matrix $S$ bringing CM (\ref{gammast2}) to the Williamson normal form (\ref{Williamson}), do not determine the matrix uniquely.
The structure of the equations reveals, that they remain unchanged under the following transformation:
\begin{equation}\label{signchange1}
x_{1}\rightarrow -x_{1},\quad x_{2}\rightarrow -x_{2},\quad x_{3}\rightarrow -x_{3},\quad x_{4}\rightarrow -x_{4},
\end{equation}
as well as under the transformation
\begin{equation}\label{signchange2}
x_{5}\rightarrow -x_{5},\quad x_{6}\rightarrow -x_{6},\quad x_{7}\rightarrow -x_{7},\quad x_{8}\rightarrow -x_{8}.
\end{equation}
Thus, by solving Eqs.~(\ref{constraints1}) and (\ref{constraints2}) we not only get the matrix
$S_{j}$, $j=1,2a,2b,3a,3b$, but also matrices
\begin{equation}\label{Ssignchange}
[\openone\oplus(-\openone)]S_{j},\quad [(-\openone)\oplus\openone]S_{j},\quad [(-\openone)\oplus(-\openone)]S_{j}.
\end{equation}
Let us stress here, that the ambiguity in determination of the matrix $S$ does not cause any nonuniqueness in determination of GIE.
This is because as we have shown in Sec.~\ref{sec_GIE}, GIE is invariant under the transformation $S\rightarrow (O_{A}\oplus O_{B})S$, where $O_{A}$
and $O_{B}$ are local orthogonal symplectic matrices and thus any of the matrices (\ref{Ssignchange}) yields the same value of GIE as the matrix
$S_{j}$. Therefore, for evaluation of GIE we can take either the matrix $S_{j}$ or any of the matrices (\ref{Ssignchange}) and for the
sake of simplicity we work with the most simple matrix $S_{j}$ in the main text.

\section{Saturation of the upper bound for states $\rho_{AB}^{(1)}$ and $\rho_{AB}^{(3)}$}\label{Lcalculation}

In this section we prove that the quantity $L$, Eq.~(\ref{L}), for symmetric GLEMS
$\rho_{AB}^{(1)}$ is equal to the RHS of Eq.~(\ref{GIEsymGLEMS}) and for asymmetric squeezed thermal GLEMS $\rho_{AB}^{(3)}$
it is equal to the RHS of Eq.~(\ref{GIEasymGLEMS}).

We start with symmetric GLEMS that are described by CM (\ref{gammast2}) with $a=b$ and $\nu_{2}=\sqrt{(a-k_{x})(a+k_{p})}=1$, Eq.~(\ref{nu12sym}).
From Eq.~(\ref{S2}) of Appendix~\ref{Sdiagonalization} it follows that $x_{3}=x_{4}=z_{A}/\sqrt{2}$, where the
parameter $z_{A}$ is defined below the equation. Inserting from here for $x_{3}$ and $x_{4}$ into the RHS of equation
(\ref{alphaAB}) further yields
\begin{eqnarray}\label{alphaclass1}
\alpha_{A}=\alpha_{B}=\left(\frac{\nu^{2}-1}{a}\right)\frac{z_{A}^{2}}{2},\quad
\alpha_{AB}=\frac{{\nu}^{2}-1}{\nu},\nonumber\\
\end{eqnarray}
and the quantity (\ref{K2}) which is to be minimized on the interval $Q\in(0,1/\nu)$ to get the quantity
(\ref{calKmin}) attains the form:
\begin{eqnarray}\label{K2class1}
\mathscr{K}_{h}&=&\frac{(1-\alpha_{A} Q)^{2}}{1-\alpha_{AB} Q}.
\end{eqnarray}
Solution of the extremal equation $d\mathscr{K}_{h}/dQ=0$ gives two stationary points
$Q_{1}=1/\alpha_{A}$ and $Q_{2}=2/\alpha_{AB}-1/\alpha_{A}$. It is easy to show with the help of
equations $\nu=\sqrt{(a+k_{x})(a-k_{p})}$ and $\sqrt{(a-k_{x})(a+k_{p})}=1$, and inequalities $k_{x}\geq k_{p}$ and $a>k_{x}$,
where the second inequality follows from the second of inequalities (\ref{CMcondition}), that $Q_{1}=1/\alpha_{A}>1/\nu$ and
thus $Q_{1}\notin(0,1/\nu)$. Likewise, using inequalities $a>k_{x}$ and $k_{p}>0$
we also find that $Q_{2}=k_{x}(a-k_{p})/[a\nu(k_{x}-k_{p})]>1/\nu$, whence $Q_{2}\notin(0,1/\nu)$. At the boundary
points $0$ and $1/\nu$ the quantity (\ref{K2class1}) then satisfies inequality $\mathscr{K}_{h}(0)=1\geq(a^2-k_{x}^{2})/(a^2-k_{p}^2)=\mathscr{K}_{h}(1/\nu)$.
As a consequence, the quantity $\mathcal{K}_{\rm min}$, Eq.~(\ref{calKmin}), is equal to $\mathcal{K}_{\rm min}=(a^2-k_{x}^{2})/(a^2-k_{p}^2)$.
Because the optimum is reached at point $1/\nu$ which has been shown below Eq.~(\ref{Kh4min}) to be reached by homodyne detection of
quadrature $x_{E}$, quantities $\mathcal{K}_{\rm min}$ and $\mathscr{K}_{\rm min}$, Eq.~(\ref{Kmin}), are equal, and one finds the quantity
(\ref{L}) to be
\begin{equation}\label{LGLEMS1}
L\left(\rho_{AB}^{(1)}\right)=\ln\left(\frac{a}{\sqrt{a^{2}-k_{p}^{2}}}\right),
\end{equation}
which clearly coincides with the RHS of Eq.~(\ref{GIEsymGLEMS}) as we set out to prove.

It is also possible to prove equality of the quantity $L$, Eq.~(\ref{L}), to GIE for asymmetric two-mode
squeezed thermal GLEMS $\rho_{AB}^{(3)}$, Eq.~(\ref{GIEasymGLEMS}). The latter states are defined by
conditions $k_{x}=k_{p}\equiv k$ and $\nu_{2}=[\sqrt{(a+b)^2-4k^2}-|a-b|]/2=1$, which lead
to the following CM \cite{Mista_16}:
\begin{eqnarray}\label{CMasymGLEMS}
\gamma_{AB}^{(3)}=\left(\begin{array}{cc}
a\openone & k\sigma_{z}\\
k\sigma_{z} & b\openone\\
\end{array}\right)
\end{eqnarray}
with
\begin{equation}\label{k}
k=\left\{\begin{array}{lll} \sqrt{(a+1)(b-1)}, & \textrm{if} & a\geq b;\\
\sqrt{(a-1)(b+1)}, & \textrm{if}  & a<b.
\end{array}\right.
\end{equation}

First, we investigate states ($\equiv\bar{\rho}_{AB}^{(3)}$) with $a\geq b$. For $a=b$ asymmetric squeezed thermal
GLEMS reduce to pure states $\rho_{AB}^{(p)}$ with $k=\sqrt{a^2-1}$, and for them the purifying subsystem
$E$ is completely decoupled from modes $A$ and $B$. This implies, that matrices (\ref{XAB})
and (\ref{Xj}) read as $X_{A}=X_{B}=X_{AB}=\gamma_{E}$ and thus $\mathscr{K}=1$ by Eq.~(\ref{K1}).
As a result, the quantity $L$ is given only by the first term on the RHS of Eq.~(\ref{L}), i.e., $L(\rho_{AB}^{(p)})=\ln(a)$, which is
equal to GIE given in Eq.~(\ref{GIEpure}). For the case $a>b$ the symplectic matrix
which symplectically diagonalizes CM (\ref{CMasymGLEMS})
is of the form (\ref{S}), where parameters $x_{1},x_{2},\ldots,x_{8}$,
are given in Eq.~(\ref{2b}). Making use of the first branch of Eq.~(\ref{k}), the explicit
form of the larger symplectic eigenvalue $\nu\equiv\nu_{1}=1+a-b$, and condition $\nu_{2}=1$, we arrive at
the parameters $x_{3}$ and $x_{4}$ appearing in Eq.~(\ref{alphaAB}),
\begin{equation}\label{x34class3a}
x_{3}=\sqrt{\frac{a+1}{a-b+2}},\quad x_{4}=\sqrt{\frac{b-1}{a-b+2}},
\end{equation}
which further give
\begin{eqnarray}\label{alphaclass3a}
\alpha_{A}&=&\frac{(a-b)(a+1)}{a},\quad \alpha_{B}=\frac{(a-b)(b-1)}{b},\nonumber\\
\alpha_{AB}&=&\frac{(a-b)(a-b+2)}{a-b+1}.
\end{eqnarray}
In this case, quantity (\ref{K2}) has two stationary points (\ref{Q12}) which can be brought
using Eq.~(\ref{alphaclass3a}) into the form $Q_{1,2}=1/(\nu\mp1)$. As $Q_{1}>1/\nu$ we have
$Q_{1}\notin(0,1/\nu)$, whereas it is obvious, that $Q_{2}\in(0,1/\nu)$. From Eqs.~(\ref{alphaclass3a})
and expression for the larger symplectic eigenvalue $\nu$ given above Eq.~(\ref{x34class3a}), it further follows
that $\mathscr{K}_{h}(1/\nu)=1$, which is equal to the value
at the other boundary point, $\mathscr{K}_{h}(0)=1$. Finally, substitution of $Q_2=1/(\nu+1)$ into the RHS
of Eq.~(\ref{K2}) and utilization of formulas (\ref{alphaclass3a}) gives
\begin{equation}\label{K2class3a}
\mathscr{K}_{h}(Q_{2})=\frac{(a+b)^{2}(a-b+1)}{ab(a-b+2)^{2}}<1.
\end{equation}
This implies, that $\mathcal{K}_{\rm min}=\mathscr{K}_{h}(Q_{2})$ and because according to Eq.~(\ref{Qfinal}) the stationary
point $Q_{2}=1/(\nu+1)$ is reached for $\varphi=0, \tau=1$ and $t=0$, which corresponds to heterodyne detection, i.e.,
projection onto a coherent state with CM $\Gamma_{E}=\openone$, $\mathcal{K}_{\rm min}$ is again equal to $\mathscr{K}_{\rm min}$.
Hence, one gets using Eq.~(\ref{L}) the expression
\begin{equation}\label{LGLEMS3a}
L\left(\bar{\rho}_{AB}^{(3)}\right)=\ln\left(\frac{a+b}{a-b+2}\right).
\end{equation}

The remaining part is a proof of the equality of GIE to the quantity $L$ for the asymmetric squeezed thermal
GLEMS with $a<b$ ($\equiv\tilde{\rho}_{AB}^{(3)}$). The proof goes along the same lines as the proof for the case with $a>b$.
Like previously, the matrix which symplectically diagonalizes CM (\ref{CMasymGLEMS}) with $k$ given by the
second branch in Eq.~(\ref{k}) is of the form (\ref{S3b}), where parameters $x_{1},x_{2},\ldots,x_{8}$,
are given in Eq.~(\ref{tildex}). This gives us for the parameters $x_{3}$ and $x_{4}$ appearing in
Eqs.~(\ref{alphaAB}) expressions
\begin{equation}\label{x34class3b}
x_{3}=\sqrt{\frac{a-1}{b-a+2}},\quad x_{4}=\sqrt{\frac{b+1}{b-a+2}}
\end{equation}
and thus
\begin{eqnarray}\label{alphaclass3b}
\alpha_{A}&=&\frac{(b-a)(a-1)}{a},\quad \alpha_{B}=\frac{(b-a)(b+1)}{b},\nonumber\\
\alpha_{AB}&=&\frac{(b-a)(b-a+2)}{b-a+1}.
\end{eqnarray}
By calculating ratios $\alpha_{AB}/\alpha_{A}$ and $\alpha_{AB}/\alpha_{B}$ and substituting the
obtained expressions into Eq.~(\ref{Q12}), we get the stationary points of the function
(\ref{K2}), $Q_{1,2}=1/(\nu\mp1)$, where now $\nu=1+b-a$. Like before, we have
$Q_{1}\notin(0,1/\nu)$, whereas $Q_{2}\in(0,1/\nu)$. Similarly, it holds that
$\mathscr{K}_{h}(0)=\mathscr{K}_{h}(1/\nu)=1$ and
\begin{equation}\label{K2class3b}
\mathscr{K}_{h}(Q_{2})=\frac{(a+b)^{2}(b-a+1)}{ab(b-a+2)^{2}}<1,
\end{equation}
whence $\mathcal{K}_{\rm min}=\mathscr{K}_{h}(Q_{2})$. Finally, since $Q_{2}$ is reached by
heterodyne detection on mode $E$ we have that $\mathcal{K}_{\rm min}=\mathscr{K}_{\rm min}$ and thus
\begin{equation}\label{LGLEMS3b}
L\left(\tilde{\rho}_{AB}^{(3)}\right)=\ln\left(\frac{a+b}{b-a+2}\right).
\end{equation}
Combining Eqs.~(\ref{LGLEMS3a}) and (\ref{LGLEMS3b}) we arrive at a single formula for quantity (\ref{L})
for all states $\rho_{AB}^{(3)}$,
\begin{equation}\label{LGLEMS3ab}
L\left(\rho_{AB}^{(3)}\right)=\ln\left(\frac{a+b}{|a-b|+2}\right).
\end{equation}
If we now compare the latter equation with Eq.~(\ref{GIEasymGLEMS}) it is clear, that also
in the case of states $\rho_{AB}^{(3)}$ the quantity $L$ coincides with the GIE which accomplishes
our proof.



\begin{thebibliography}{99}
\bibitem{PES} In public Eve scenario \cite{Christandl_04}, Eve
must choose a function, which is applied to $E$ to obtain
$\tilde{E}$. The value of $\tilde{E}$ together with the
description of the function is then broadcasted to Alice and Bob.
\end{thebibliography}
\end{document}